\documentclass[fleqn,usenatbib]{mnras}

\usepackage{newtxtext,newtxmath}
\usepackage{txfonts}

\usepackage[T1]{fontenc}

\DeclareRobustCommand{\VAN}[3]{#2}
\let\VANthebibliography\thebibliography
\def\thebibliography{\DeclareRobustCommand{\VAN}[3]{##3}\VANthebibliography}

\usepackage{graphicx}	
\usepackage{amsmath}	
\usepackage{amssymb}	
\usepackage{multirow}
\usepackage{adjustbox}
\usepackage{comment}
\usepackage{tablefootnote}
\usepackage{ulem}

\newcommand{\pkg}[1]{\textsc{\texttt{#1}}}

\newcommand{\orcid}[1]{\href{https://orcid.org/#1}{\includegraphics[height=11pt]{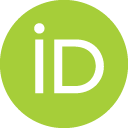}}}

\newcommand{\pysedm}{\pkg{pysedm}}
\newcommand{\byecr}{\pkg{byecr}}
\newcommand{\contsep}{\pkg{contsep}}

\newcommand{\byecontex}{\textit{byecont}}

\newcommand{\snid}{\pkg{SNID}}
\newcommand{\superfit}{\pkg{SuperFit}}
\newcommand{\ngsf}{\pkg{NGSF}}
\newcommand{\dash}{\pkg{DASH}}
\newcommand{\sniascore}{\texttt{SNIascore}}
\newcommand{\gelato}{\pkg{GELATO}}

\newcommand{\rrlap}{\textit{rlap}}
\newcommand{\chidof}{\textit{chi2/dof}}
\newcommand{\prob}{\textit{Probability}}
\newcommand{\score}{\textit{score}}

\newcommand{\qof}{\textit{QoF}}

\newcommand{\snr}{\textit{S/N}}

\defcitealias{Kim2022}{K22}

\title[Accuracy of transient spectral classification tools]{How accurate are transient spectral classification tools?\\ - A study using 4,646 SEDMachine spectra}

\author[Young-Lo Kim et al.]{
	Young-Lo Kim$^{1}$\thanks{E-mail: y.kim9@lancaster.ac.uk}\orcid{0000-0002-1031-0796}, 
	Isobel Hook$^{1}$\orcid{0000-0002-2960-978X}, 
	Andrew Milligan$^{1}$\orcid{0009-0006-6426-2431}, 
	Llu\'is Galbany$^{2,3}$\orcid{0000-0002-1296-6887},
	Jesper Sollerman$^{4}$\orcid{0000-0003-1546-6615},
	\newauthor
	Umut Burgaz$^{5}$\orcid{0000-0003-0126-3999},
	Georgios Dimitriadis$^{5}$\orcid{0000-0001-9494-179X},
	Christoffer Fremling$^{6,7}$\orcid{0000-0002-4223-103X},
	Joel Johansson$^{8}$\orcid{0000-0001-5975-290X},
	\newauthor
	Tom\'as E. M\"uller-Bravo$^{2,3}$\orcid{0000-0003-3939-7167},
	James D. Neill$^{7}$\orcid{0000-0002-0466-1119},
	Jakob Nordin$^{9}$\orcid{0000-0001-8342-6274},
	Peter Nugent$^{10,11}$\orcid{0000-0002-3389-0586},
	Josiah Purdum$^{6}$\orcid{0000-0003-1227-3738},
	\newauthor
	Yu-Jing Qin$^{7}$\orcid{0000-0003-3658-6026},
	Philippe Rosnet$^{12}$\orcid{0000-0002-6099-7565},
	Yashvi Sharma$^{7}$\orcid{0000-0003-4531-1745}
\\
$^{1}$Department of Physics, Lancaster University, Lancs LA1 4YB, UK \\
$^{2}$Institute of SpaceSciences (ICE,CSIC), Campus UAB, Carrer de Can Magrans, s/n, E-08193 Barcelona, Spain \\
$^{3}$Institut d’Estudis Espacials de Catalunya (IEEC), E-08034 Barcelona, Spain \\
$^{4}$The Oskar Klein Centre, Department of Astronomy, Stockholm University, AlbaNova, SE-106 91 Stockholm, Sweden \\   
$^{5}$School of Physics, Trinity College Dublin, College Green, Dublin 2, Ireland \\
$^{6}$Caltech Optical Observatories, California Institute of Technology, Pasadena, CA 91125, USA \\
$^{7}$Division of Physics, Mathematics and Astronomy, California Institute of Technology, Pasadena, CA 91125, USA \\
$^{8}$The Oskar Klein Centre, Department of Physics, Stockholm University, AlbaNova, SE-106 91 Stockholm, Sweden \\  
$^{9}$Institut für Physik, Humboldt-Universität zu Berlin, Newtonstr. 15, 12489 Berlin, Germany \\
$^{10}$Lawrence Berkeley National Laboratory, 1 Cyclotron Road MS 50B-4206, Berkeley, CA, 94720, USA \\
$^{11}$Department of Astronomy, University of California, Berkeley, 501 Campbell Hall, Berkeley, CA 94720, USA \\
$^{12}$Université Clermont Auvergne, CNRS/IN2P3, LPCA, F-63000 Clermont-Ferrand, France \\
}

\date{Accepted XXX. Received YYY; in original form ZZZ}

\pubyear{2024}

\begin{document}
\label{firstpage}
\pagerange{\pageref{firstpage}--\pageref{lastpage}}
\maketitle

\begin{abstract}
Accurate classification of transients obtained from spectroscopic data are important to understand their nature and discover new classes of astronomical objects.
For supernovae (SNe), \snid{}, \ngsf{} (a Python version of \superfit{}), and \dash{} are widely used in the community.
Each tool provides its own metric to help determine classification, such as \rrlap{} of \snid{}, \chidof{} of \ngsf{}, and \prob{} of \dash{}.
However, we do not know how accurate these tools are, and they have not been tested with a large homogeneous dataset.
Thus, in this work, we study the accuracy of these spectral classification tools using 4,646 SEDMachine spectra, which have accurate classifications obtained from the Zwicky Transient Facility Bright Transient Survey (BTS).
Comparing our classifications with those from BTS, we have tested the classification accuracy in various ways.
We find that \ngsf{} has the best performance (overall \textit{Accuracy} 87.6\% when samples are split into SNe Ia and Non-Ia types), while \snid{} and \dash{} have similar performance with overall \textit{Accuracy} of 79.3\% and 76.2\%, respectively.
Specifically for SNe Ia, \snid{} can accurately classify them when \rrlap{} > 15 without contamination from other types, such as Ibc, II, SLSN, and other objects that are not SNe (\textit{Purity} > 98\%).
For other types, determining their classification is often uncertain.
We conclude that it is difficult to obtain an accurate classification from these tools alone.
This results in additional human visual inspection effort being required in order to confirm the classification.
To reduce this human visual inspection and to support the classification process for future large-scale surveys, this work provides supporting information, such as the accuracy of each tool as a function of its metric.
\end{abstract}





\section{Introduction}
\label{sec:intro}
Spectroscopy provides the most unambiguous classification of transients, and indeed the main classification schemes are based on the presence and absence of certain features that are visible in the spectrum \citep[see][for a review of transient classification]{Filippenko1997}.
From accurate classification, we can understand their nature, detect new classes of astronomical objects, and estimate unbiased cosmological parameters from Type Ia supernovae (SNe Ia) by removing contamination from non-Ia types.
In modern data-intensive time-domain astronomy, classification for the ever-increasing number of new transients is one of the main challenges.
Current surveys, such as the Zwicky Transient Facility \citep[ZTF;][]{Bellm2019, Graham2019, Masci2019, Dekany2020}, the All-Sky Automated Survey for Supernovae \citep{Shappee2014}, and the Asteroid Terrestrial Last-Alert System \citep{Tonry2018}, are discovering $O$($10^{3}$) interesting transients among $O$($10^{6}$) alerts every night.
Furthermore, the Rubin Observatory’s Legacy Survey of Space and Time (LSST) will soon detect ten times more transients than current surveys \citep{lsst2009}.
However, due to a shortage of spectroscopic follow-up resources, \citet{Kulkarni2020} reported that only about 10 per cent of transients were classified spectroscopically, based on analysis of the demographics of astronomical transients and SN classifications reported to the Transient Name Server (TNS)\footnote{\href{https://wis-tns.weizmann.ac.il/}{https://wis-tns.weizmann.ac.il/}} in 2019.

In recent years, a major effort has been underway to develop classification methods using photometric data, which are easier to obtain than spectroscopic data, based on machine-learning algorithms \citep[e.g.,][]{Moller2020, Miranda2022, Gagliano2023}.
These tools are fast and efficient, and currently reach quite high accuracy ($\sim$90\%).
However, they remain in their infancy \citep[see][for a discussion]{Gagliano2023} and need to prove their reliability with more observational data \citep[see, e.g.,][]{Fremling2021}.

On the other hand, efforts are being made to obtain more spectra of transients, which can provide accurate classification.
One example is the Spectral Energy Distribution Machine \citep[SEDM;][]{sedm}, a spectroscopic follow-up instrument dedicated to ZTF-discovered transients.
SEDM is a very low-resolution ($R\sim100$) integral field unit spectrograph with a ~28 x 28 $arcsec^{2}$ field-of view, mounted on the Palomar 60-inch telescope. 
The main aim of SEDM is to classify ZTF-discovered transients spectroscopically.
\pysedm{} is a pipeline that calibrates and extracts SEDM spectra within minutes of the exposure completion.
With this pipeline, SEDM is fully automated, and hence a flux-calibrated spectrum in the wavelength range between 3500$\AA$ and 9500$\AA$ is available within 5 minutes after the end of the exposure.
In this way, 15 transient spectra are typically obtained every night.
Since early August 2018, SEDM with \pysedm{} has been in production, obtaining several thousand spectra per year (see Tab.~\ref{tab:sample}).
For LSST discovered transients, the Time-Domain Extragalactic Survey \citep[TiDES;][]{Swann2019} conducted on the 4-metre Multi-Object Spectrograph Telescope \citep[4MOST;][]{deJong2019} plans to perform spectroscopic follow-up observations, expecting to obtain spectra of more than 35,000 live transients (C. Frohmaier et al., in preparation).

Spectra of transients are usually classified through spectral classification tools.
Based on the template matching technique, the Supernova Identification code \citep[\snid{};][]{Blondin2007}, \superfit{} \citep{Howell2005}, and the GEneric cLAssification TOol \citep[\gelato{};][]{gelato} are widely used in the community.
In addition, there are deep-learning-based tools that are often utilized: Deep Automated Supernova and Host classifier \citep[\dash{};][]{Muthukrishna2019} and \sniascore{} \citep{Fremling2021}.
Each tool has its own metric to help determine classification: \rrlap{} of \snid{}, \chidof{} of \superfit{}, \qof{} of \gelato{}, \prob{} of \dash{}, and \score{} of \sniascore{}.
Even though those tools are widely adopted in the community for the spectroscopic classifications of transients, their performance has not been studied based on confusion matrix and performance evaluation metrics, except for \sniascore{}, which was specifically designed for spectroscopic classification for SNe Ia discovered by ZTF.
Thus, we do not know how accurate those tools and their metrics are.
This might be due to the fact that there has not been a sufficient number of accurately classified and homogeneous transient spectroscopic data to perform quantitative studies thus far.
If their performance is tested with such data and from this, if we can provide supporting information for spectroscopic classification (e.g., the accuracy of each tool as a function of its metric), it would be helpful for future spectroscopic follow-up surveys that will observe an order of magnitude more transients than current surveys.

The Bright Transient Survey \citep[BTS; ][]{BTS, Perley2020}, one of the ZTF sub-surveys, has the primary goal of spectroscopically classifying and publicly reporting every extragalactic transient brighter than 18.5 mag in the $r$-band in the Northern sky detected by the ZTF public survey.
BTS uses transients' spectra obtained from SEDM and from various larger telescopes, e.g., the Palomar 200-inch telescope, Keck I, the Liverpool Telescope, Apache Point Observatory, and the Nordic Optical Telescope.
Preliminary classifications are made via \snid{} and human visual inspection.
Then, the BTS team revisits each preliminary classification through a homogeneous process they developed.
For each SN spectrum, the team identifies the top 15 matched templates from \snid{} and produces plots showing a comparison between the observed spectrum and the \snid{} template spectrum.
These plots are visually inspected by a member of the BTS team to identify the best matching templates.
In cases where all 15 \snid{} templates report the same classification, the \snid{} type is recorded.
Otherwise, the BTS team selects the best classification based on the SN spectral features.
If the top 15 matches prove ambiguous, they use additional information, such as the light-curve or alternative spectra.
If a classification is still uncertain, then this SN spectrum is examined by another member of the team.
For consistency, a final check is performed by two members of the team.
From this substantial additional effort, BTS provides a large number of accurate spectroscopic classifications for ZTF-detected transients.
We refer the reader to \citet{BTS} for details of the BTS program.

In this work, we use a large sample of homogeneous transients' spectra from SEDM and their classifications from BTS to test the classification accuracy of \snid{}, \superfit{}, and \dash{}\footnote{We select those three tools, because they are easily implemented with $Python$. See Sec.~\ref{subsec:tools}.}.
Then, based on this test, we provide supporting metrics for spectroscopic classifications.

\section{Method}
\label{sec:method}


\subsection{Sample}
\label{subsec:sample}

%
\begin{table*}
\centering
\caption{The sample size of SEDM spectra between August 2018 and June 2022. Aug2018-2020 data are taken from \citetalias{Kim2022}.}
\label{tab:sample}
\begin{tabular}{l c c c | c}
\hline\hline\\[-0.8em]
                             & Aug2018-2020       & 2021         & Jan-Jun2022  & Total \\[0.15em] 
\hline\\[-0.8em]
all observed spectra             &  6784               	& 2884                 & 1154            	& 10822  \\ [0.30em]
+ extraction with the \byecontex{} method           &  6656               	& 2328         	   &  942                   	& 9926 \\ [0.30em]
+ \snr{} > 3                  & 5852              	& 2120          	   & 828                     & 8800 \\ [0.30em]
+ classification \textbf{from BTS} (final sample)        & \textbf{3025}     & \textbf{1284}    & \textbf{337}		&\textbf{4646}    \\ [0.30em]
\hline
\end{tabular}
\end{table*}

Tab.~\ref{tab:sample} shows the sample size of SEDM spectra used in this work.
We begin with 10,822 SEDM spectra observed between August 2018 and June 2022.

From these, we selected 9,926 spectra extracted by the default \pysedm{} pipeline with two modules, namely \byecr{}  and \contsep{}, developed by \citet[][hereafter \citetalias{Kim2022}]{Kim2022}.
The \byecr{} module removes contamination from cosmic rays, and \contsep{} removes non-target contamination from the host galaxy and other nearby sources.
\citetalias{Kim2022} defined this combination as \byecontex{}: \pysedm{} + \byecr{}  + \contsep{}.
The \contsep{} module requires an astrometry solution from SEDM and a reference image, centred at the target coordinate, to draw isomagnitude contours.
Then, the \contsep{} module tries to find the faintest contour to separate the target from other sources.
If we have an inaccurate astrometry solution due to a SEDM pointing failure, a reference image issue (e.g., saturation), or a target that is too close to the host core, the \contsep{} module does not work.
This means the \byecontex{} method has fewer extracted spectra than the default \pysedm{} pipeline.
However, the \byecontex{} method showed improvements in the classification rate and accuracy over the default \pysedm{} pipeline (see \citetalias{Kim2022} for a discussion).
Therefore, we selected the \byecontex{} extraction method to test the accuracy of spectral classification tools.

Next, we apply a cut based on the signal-to-noise ratio (\snr{}), following the method of \citetalias{Kim2022}.
The \snr{} (per $\sim$25 $\AA$ bin) is calculated as the median \snr{} in the wavelength range between 4000 and 8000 $\AA$ that will be adopted for spectroscopic classification.
We selected all the 8,800 (out of 9,926) SEDM spectra that have \snr{} > 3.

Lastly, we need true classifications for our sample to test the classification accuracy.
This true classification is from BTS.
Matching our 8,800 spectra to the BTS sample returned 4,646 spectra (for 2,986 individual targets) with their classification and redshift ($z_{BTS}$).
We use this sample of 4,646 spectra as our final sample for testing the accuracy of spectral classification tools.

%
\begin{figure}
 \centering
  	\includegraphics[width=\columnwidth]{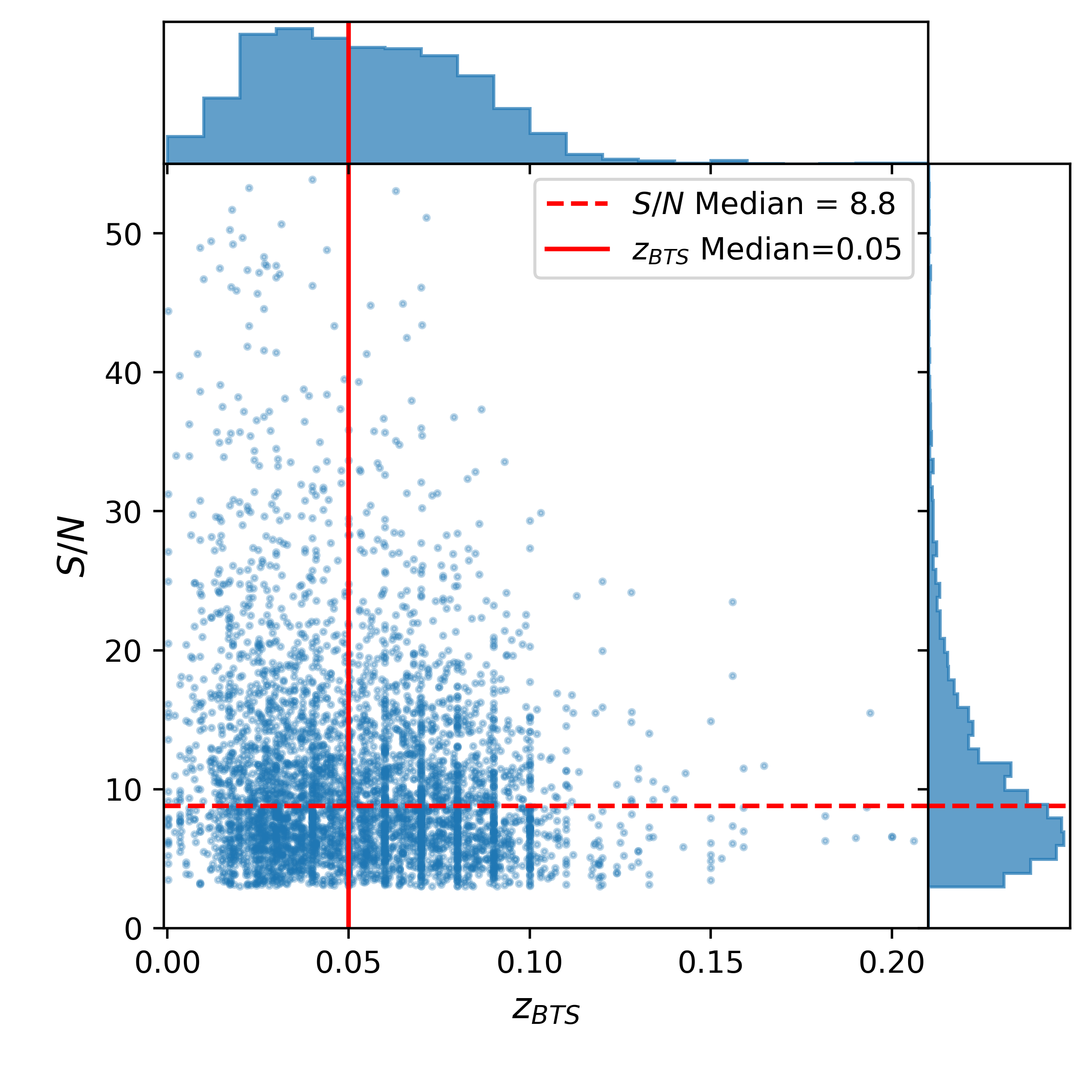}
  \caption{\snr{} and BTS redshift ($z_{BTS}$) distribution of our sample of 4,646 SEDM spectra.
  		The median of \snr{} and $z_{BTS}$ are indicated with red dotted and solid lines, respectively.
  		Note that vertical structures shown in the figure are due to redshifts estimated from transients' spectra \citep[see][for a full description of redshift measurements]{BTS}.
  		}
  \label{fig:sample_distribution}
\end{figure}

Fig.~\ref{fig:sample_distribution} shows the \snr{} and $z_{BTS}$ distributions of our final sample of 4,646 SEDM spectra.
The median $z_{BTS}$ is 0.05, and the median \snr{} is 8.8.
In the figure, some vertical structures are shown.
We found that this is due to redshifts that were estimated from transients' spectra with \snid{}. That method provided up to two significant figures, whereas redshift from host spectra returned more significant figures.
The lower number of significant figures for some spectra results in vertical structures in the figure.
However, \citet{BTS} discussed that the redshift of the transients is a good estimator of the host galaxy redshift \citep[see][for a full description of redshift measurements]{BTS}.
We note that since we do not use $z_{BTS}$ when running the spectral classification tools below, this does not affect our analysis.


\subsection{Spectral classification tools: \snid{}, \ngsf{}, and \dash{}}
\label{subsec:tools}

This work will test the classification accuracy of three widely used spectral classification tools: \snid{}\footnote{\href{https://people.lam.fr/blondin.stephane/index.html}{https://people.lam.fr/blondin.stephane/index.html}}, \superfit{}, and \dash{}\footnote{\href{https://github.com/daniel-muthukrishna/astrodash}{https://github.com/daniel-muthukrishna/astrodash}}.
For \snid{}, we use its $Python$ wrapper \texttt{pysnid}\footnote{\href{https://github.com/MickaelRigault/pysnid}{https://github.com/MickaelRigault/pysnid}}, and for \superfit{}, we use its $Python$ version, \ngsf{}\footnote{\href{https://github.com/oyaron/NGSF}{https://github.com/oyaron/NGSF}} \citep[Next Generation \superfit{};][]{ngsf}.

When running the spectral classification tools, we set a wavelength range parameter to use from 4000 and 8000 $\AA$, where the quantum efficiency of the SEDM CCD is over 60 per cent and which includes most of the important spectral features of the transients, as in \citetalias{Kim2022}.
We employed default input values provided by each tool, except for values for the extinction law in \ngsf{}, for which we use \textit{`Alam\_high' : 3} and  \textit{`Alam\_low' : --3}, instead of the default values of \textit{`Alam\_high' : 2} and  \textit{`Alam\_low' : --2}.
We put a redshift range between 0.0 and 0.2, considering the redshift range of ZTF\footnote{In this work, we do not use redshift (e.g., $z_{BTS}$) when running the tools to reflect the actual observation situation, such as SEDM working with its \pysedm{} pipeline.} (e.g., see Fig.~\ref{fig:sample_distribution} for BTS).

%
\begin{table*}
\centering
\caption{Types and sub-types in each category.}
\label{tab:types}
\resizebox{\textwidth}{!}{%
\begin{tabular}{l c c l}
\hline\hline\\[-0.8em]
                             & 2 Categories & 5 Categories & (sub-types)  \\[0.15em] 
\hline\\[-0.8em]
\snid{}   & Ia        & Ia           &  (Ia,Ia-norm,Ia-91T,Ia-91bg,Ia-csm,Ia-pec,Ia-99aa,Ia-02cx,SN Ia,SN Ia-91T,SN Ia-91bg,SN Ia-CSM,SN Ia-pec,SN Iax)   \\ [0.30em]
             & NonIa  & Ibc          & (Ib,Ib-norm,Ib-pec,IIb,Ibn,SN Ib,SN Ib-pec,SN Ib/c,SN Ibn,SN IIb,Ic,Ic-norm,Ic-pec,Ic-broad,SN Ic,SN Ic-BL,SN Icn,Ca-rich)         \\ [0.30em]
             &		 & II             &  (II,IIP,II-pec,IIn,IIL,SN II,SN II-pec,SN IIP,SN IIn)     \\ [0.30em]
             &            & SLSN      & (SLSN,SLSN-I,SLSN-Ic,SLSN-IIn,SLSN-II)            \\ [0.30em]
             &            & NotSN     & (NotSN,AGN,Gal,LBV,M-star,QSO,C-star,TDE,-,ILRT,LRN,Other,nova,other)              \\ [0.50em]
\ngsf{}   & Ia        & Ia           &  (Ia 02es-like,Ia 91T-like,Ia 91bg-like,Ia 99aa-like,Ia-02cx like,Ia-CSM,Ia-CSM-(ambigious),Ia-norm,Ia-pec,Ia-rapid,Ca-Ia,super chandra)   \\ [0.30em]
             & NonIa  & Ibc          & (Ib,Ibn,IIb,IIb-flash,Ic,Ic-BL,Ic-pec,Ca-Ib)         \\ [0.30em]
             & 		 & II             &  (II,II-flash,IIn)     \\ [0.30em]
             &            & SLSN      & (SLSN-I,SLSN-II,SLSN-IIb,SLSN-IIn,SLSN-Ib)            \\ [0.30em]
             &            & NotSN     & (FBOT,ILRT,SN - Imposter,TDE H,TDE H+He,TDE He)              \\ [0.50em]
\dash{}  & Ia        & Ia           &  (Ia,Ia-02cx,Ia-91T,Ia-91bg,Ia-csm,Ia-norm,Ia-pec)   \\ [0.30em]
             & NonIa  & Ibc          & (Ib,Ib-norm,Ib-pec,Ibn,IIb,Ic,Ic-broad,Ic-norm,Ic-pec)         \\ [0.30em]
             & 		& II             &  (II,II-pec,IIP,IIn,IIL)     \\ [0.50em]
BTS      & Ia        & Ia           &  (SN Ia,SN Ia-91T,SN Ia-91bg,SN Ia-CSM,SN Ia-pec,SN Iax,SN Ia-SC)   \\ [0.30em]
             & NonIa  & Ibc          & (SN Ib,SN Ib-pec,SN Ib/c,SN Ibn,SN Ic,SN Ic-BL,SN Icn,SN IIb,Ca-rich)         \\ [0.30em]
             & 		  & II             &  (SN II,SN IIP,SN IIn,SN II-pec)     \\ [0.30em]
             &            & SLSN      & (SLSN-I,SLSN-II)            \\ [0.30em]
             &            & NotSN     & (ILRT,LBV,TDE,nova,other,Other)              \\ [0.50em]
\hline
\end{tabular}}
\end{table*}

Regarding the templates, we used the default templates for \ngsf{} and \dash{}, while \snid{} templates were taken from \citetalias{Kim2022}.
The templates used are collected from a training set for \dash{} (which combines the spectra from \snid{} Templates 2.0, the Berkeley SN Ia program v7.0 \citep{Silverman2012}, SN Ib/c from \citet{Liu2014}, \citet{Liu2016}, and \citet{Modjaz2016}), as well as SN IIP templates from \citet{Gutierrez2017}, SLSN-Ic from \citet{Liu2017}, and several SLSN-I, SLSN-IIn and TDE added by J.D.Neill.

Based on the types and sub-types in the templates of each tool, we create broad categories for the transients: `2 Categories' for `Ia' and `NonIa', and `5 Categories' for `Ia', `II', `Ibc', `SLSN', and `NotSN' (Tab.~\ref{tab:types}).
This category scheme will be adopted when testing the accuracy of spectral classifications in this work.
We note that \dash{} does not have `SLSN' and `NotSN' categories.

\section{Run results of \snid{}, \ngsf{}, and \dash{}}
\label{sec:results}

%
\begin{table}
\centering
\caption{The size of successfully classified samples for each classification tool split by our category scheme.}
\label{tab:run_results}
\begin{tabular}{l r r r r}
\hline\hline\\[-0.8em]
                                     & \snid{}            & \ngsf{}            & \dash{}    & BTS \\[0.15em] 
\hline\\[-0.8em]
\textbf{Total}                 & \textbf{4538} & \textbf{4646} & \textbf{4646}  &     \textbf{4646} \\
\hline
SN Ia                             & 2798            & 2561            & 2446          & 2831 \\ [0.30em]
NonIa                            & 1740            &  2085            & 2200          & 1815 \\ 
\hline
 \hspace{4mm}SN Ibc   & 1014            & 660               & 1784         & 596 \\ [0.30em]
 \hspace{4mm}SN II      & 354              & 931               & 416           & 1069 \\ [0.30em]
 \hspace{4mm}SLSN    & 142               & 408               & -               & 95 \\ [0.30em]
 \hspace{4mm}NotSN  & 230               & 86                 & -               & 55 \\ 
\hline
(Failed Classification                 & 108               & 0                    & 0              & - ) \\
\hline
\end{tabular}
\end{table}

%
\begin{figure*}
 \centering
  	\includegraphics[width=0.45\textwidth]{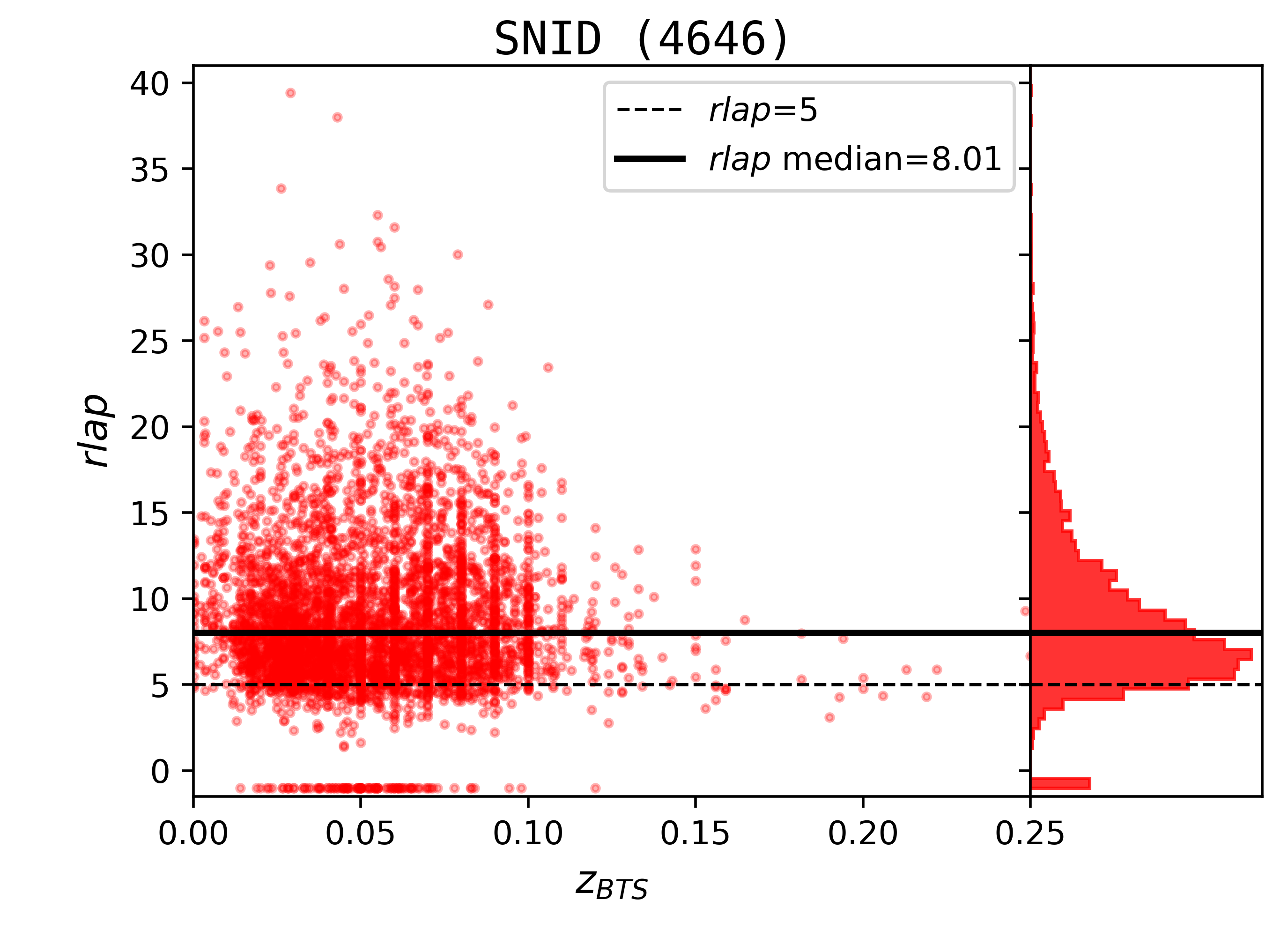}
	\includegraphics[width=0.45\textwidth]{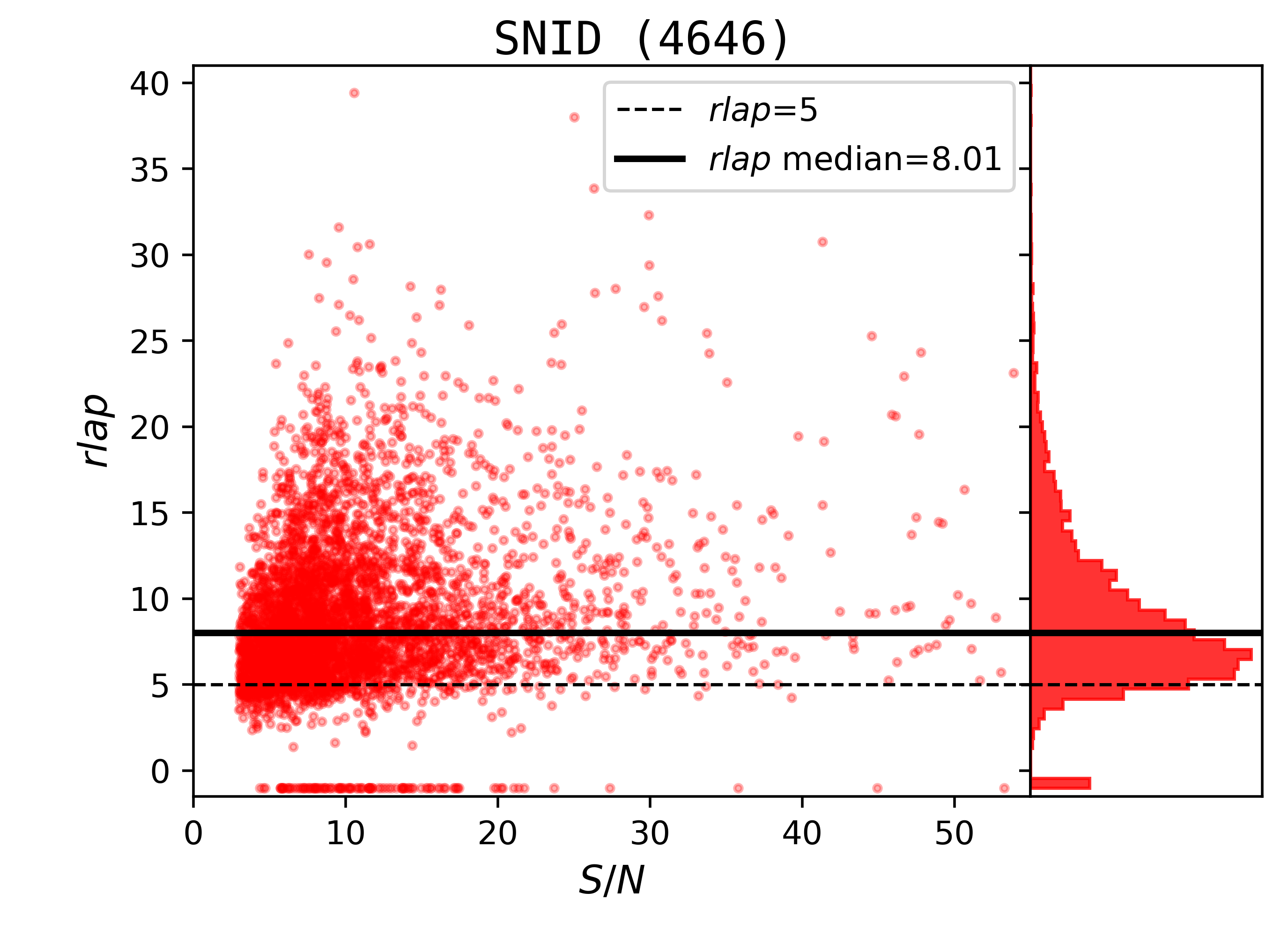}
	\includegraphics[width=0.45\textwidth]{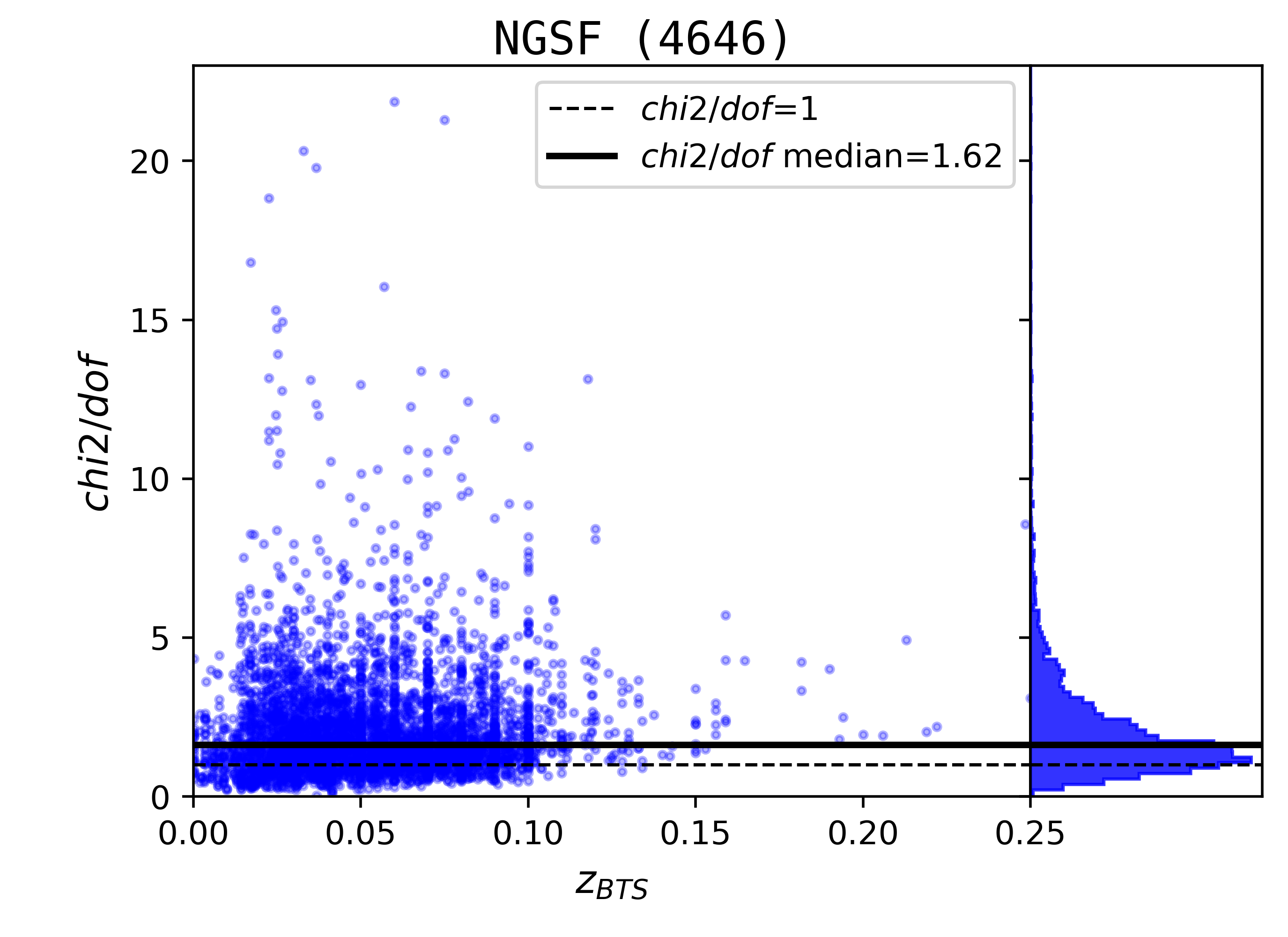}
	\includegraphics[width=0.45\textwidth]{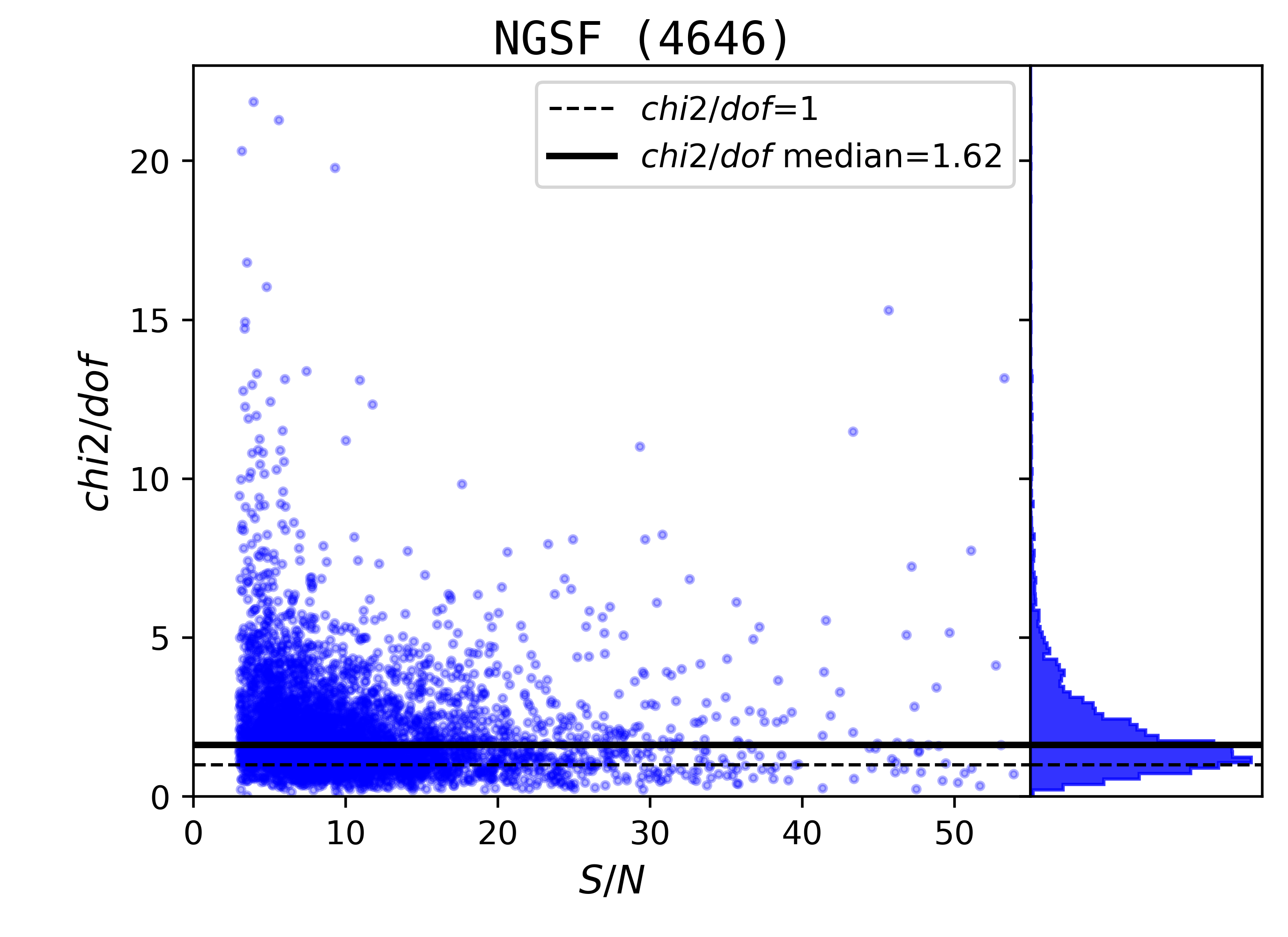}
	\includegraphics[width=0.45\textwidth]{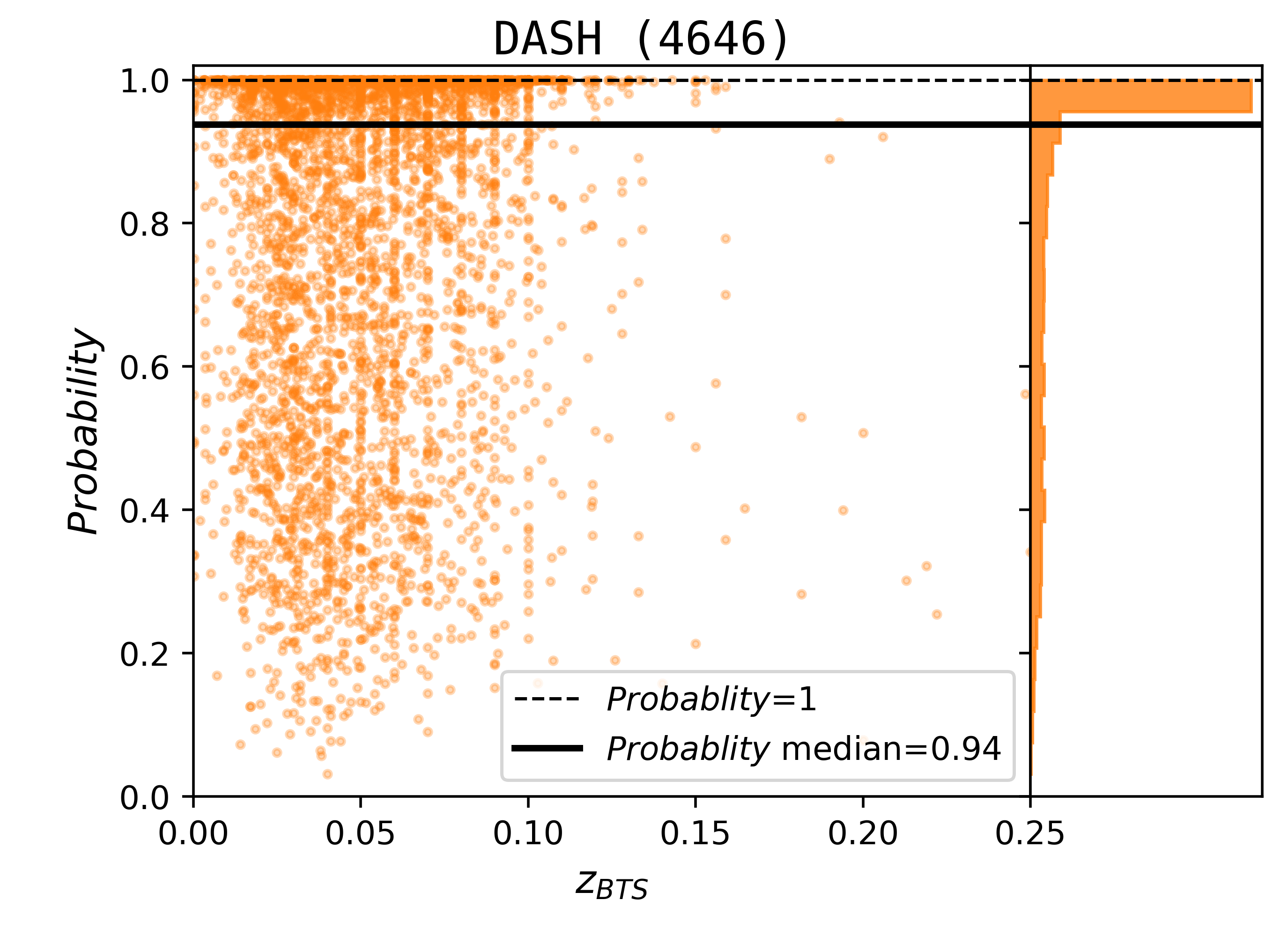}
	\includegraphics[width=0.45\textwidth]{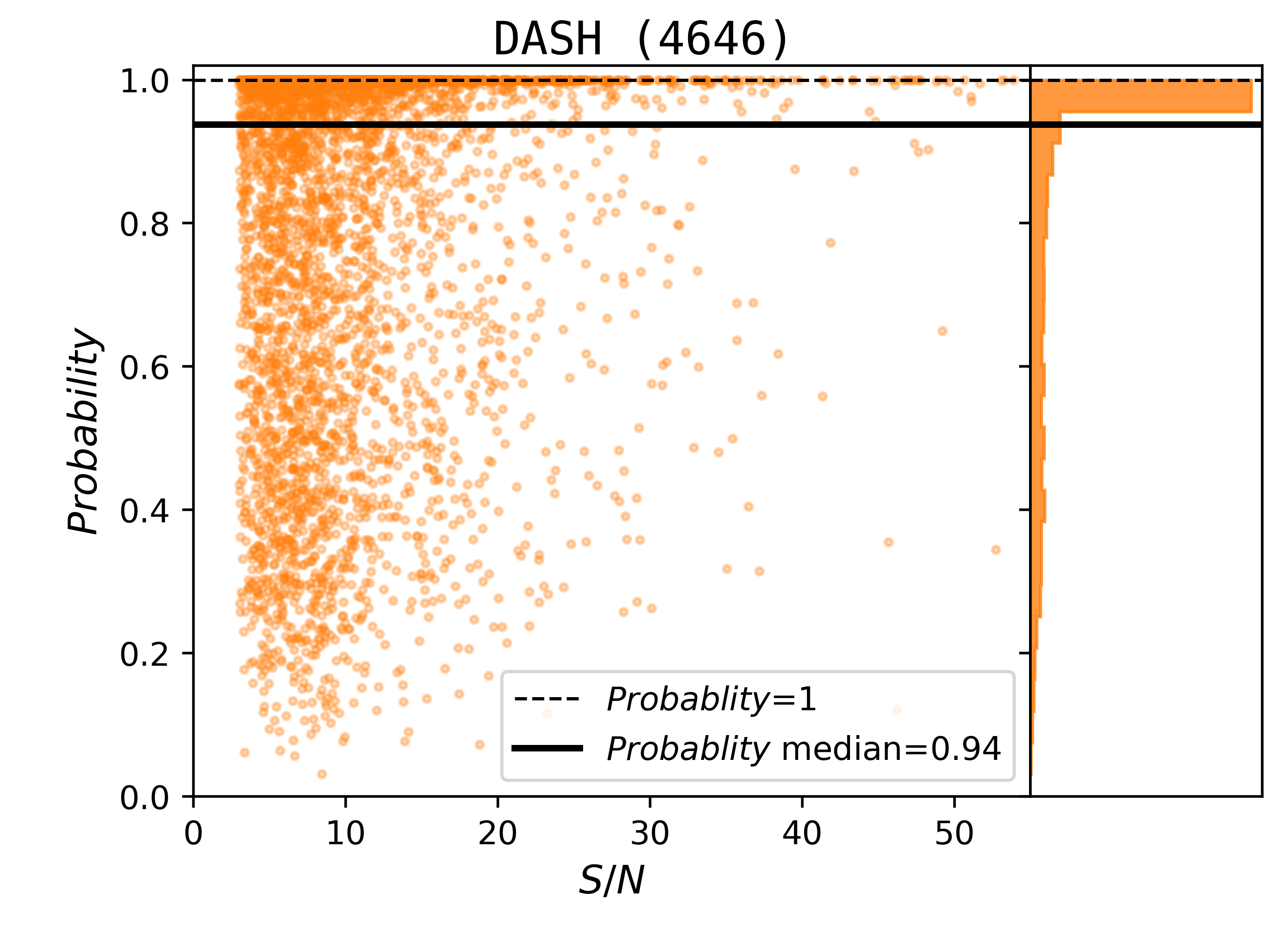}
  \caption{Distributions of each tool's primary metric as a function of $z_{BTS}$ (left panel) and the \snr{} of the SEDM spectrum (right).
  		Solid lines indicate the median of each metric, and dashed lines represent the classification quality criteria.
		For \snid{}, \rrlap{} = 5 is the minimum value for a reliable classification, as suggested by \citet{Blondin2007}.
		For \ngsf{} and \dash{}, \chidof{} = 1.0 and \prob{} = 1.0 mean the best-fitted and relatively more likely classification with a specific template, respectively.
		108 failed \snid{} classifications (3.9\%) are presented with \rrlap{} = --1 (top panel).
		High \snr{} spectra appear to converge around \chidof{} = 1.0 and \prob{} = 1.0, while high \snr{} does not guarantee high \rrlap{} values.
		No trends are found with $z_{BTS}$.
		We note that we use different colours to easily distinguish each tool: red for \snid{}, blue for \ngsf{}, and orange for \dash{}.
  		}
  \label{fig:quality_value_distribution}
\end{figure*}

Tab.~\ref{tab:run_results} shows the output sample size of each classification tool split by our category scheme after running \snid{}, \ngsf{}, and \dash{}.
There are 108 failed cases (3.9\%) where \snid{} could not find a correlation.
We flagged these cases as \rrlap{} = --1 for further inspection (see the top panel of Fig.~\ref{fig:quality_value_distribution}).

For the analysis below, we employ the primary metric for each tool: \rrlap{} of \snid{}, \chidof{} of \ngsf{}, and \prob{} of \dash{}.
The \snid{} \rrlap{} value is a measurement of the strength of the correlation between a template spectrum and the input spectrum.
Thus, a higher \rrlap{} value means a more secure classification.
\chidof{} of \ngsf{} is a useful measure of goodness of fit.
\ngsf{} returns the best-fitted classification based on finding the minimum $\chi^{2}$ value by comparing the observed spectrum to the library of templates.
During this $\chi^{2}$ fitting process, \ngsf{} tries to separate the contributions to the spectrum from the transient and its host galaxy by fitting them simultaneously.
\dash{} \prob{} provides a relative probability of a particular classification compared to other types.
Thus, a higher value of \prob{} means a specific classification is more likely than other types.
Considering the definition of each primary metric, we select a classification with the highest \rrlap{} value for \snid{}, the lowest \chidof{} for \ngsf{}, and the highest \prob{} for \dash{} as our classification in the analysis below.

In Fig.~\ref{fig:quality_value_distribution}, we plot distributions of each primary metric as a function of $z_{BTS}$ and the \snr{} of the SEDM spectrum.
Solid lines indicate the median of each metric.
Dashed lines represent the criteria for the goodness of fit with templates, in other words, the classification quality criteria: \chidof{} = 1.0 for \ngsf{} and \prob{} = 1.0 for \dash{}.
For \snid{}, \rrlap{} = 5 for \snid{} is the minimum value for reliable classification, as suggested by \citet{Blondin2007}.
With $z_{BTS}$, no trends are found, while some trends are observed with \snr{}.
We find that high \snr{} spectra appear to converge to \chidof{} = 1.0 and \prob{} = 1.0.
However, high \snr{} does not guarantee high \rrlap{} values.
Note that in the figures below, we use different colours to easily distinguish each tool: red for \snid{}, blue for \ngsf{}, and orange for \dash{}.

%
\begin{figure*}
 \centering
  	\includegraphics[width=\textwidth]{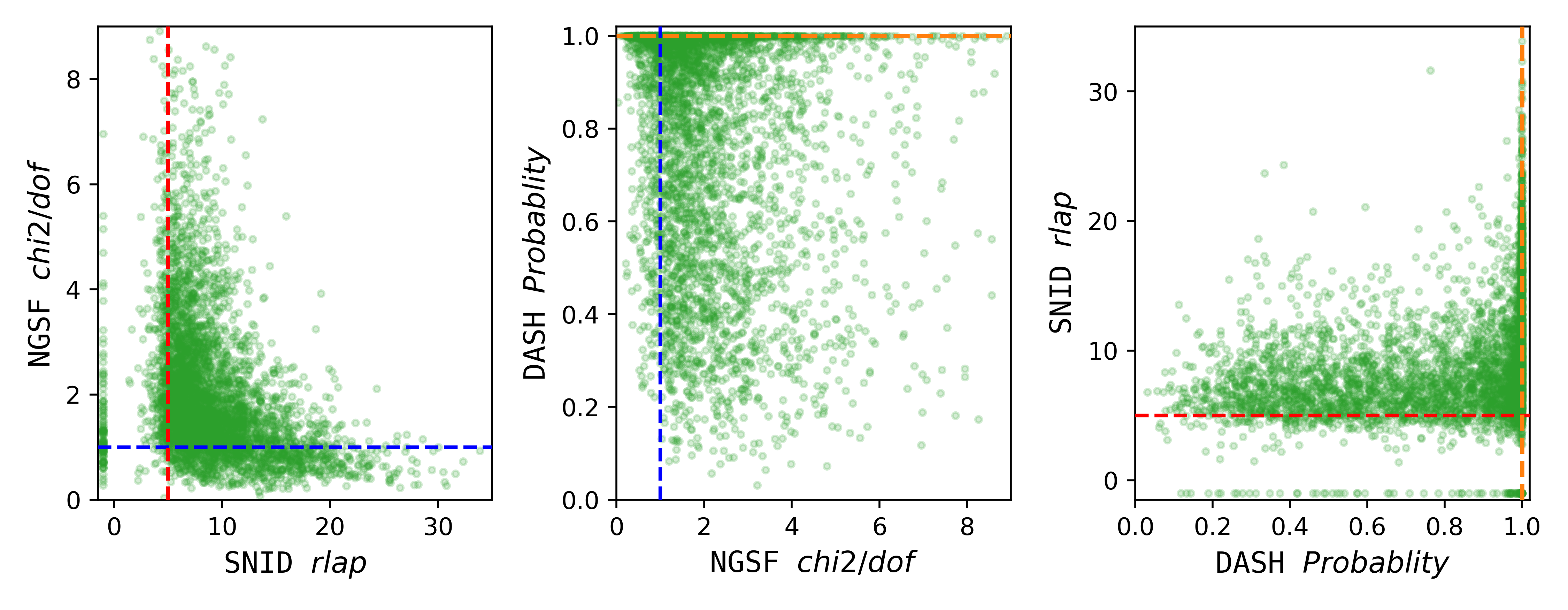}
  \caption{Correlation between the primary metrics of pairs of tools.
  		Dashed lines with different colours represent the classification quality criteria (same as dashed lines in Fig.~\ref{fig:quality_value_distribution}): \snid{} \rrlap{} = 5 (red), \ngsf{} \chidof{} = 1.0 (blue), and \dash{} \prob{} = 1.0 (orange).
		108 failed \snid{} classifications (3.9\%) are presented with \rrlap{} = --1.
		High \rrlap{} values of \snid{} tend to converge to \ngsf{} \chidof{} $\sim$ 1.0 (left panel) and \dash{} \prob{} $\sim$ 1.0 (right).
		Between \ngsf{} and \dash{} (middle panel), a cluster is formed around \chidof{} $\sim$ 1 and \prob{} $\sim$ 1.0.
  		}
  \label{fig:quality_value_comp}
\end{figure*}

Fig.~\ref{fig:quality_value_comp} shows the correlation between the primary metrics of pairs of tools.
We find that high \rrlap{} values of \snid{} tend to converge to \ngsf{} \chidof{} $\sim$ 1.0 and \dash{} \prob{} $\sim$ 1.0.
Between \ngsf{} and \dash{}, a cluster is formed around \chidof{} $\sim$ 1 and \prob{} $\sim$ 1.0.
However, there are many cases where a classification has a value of \chidof{} $\sim$ 1, but with a low \prob{} value.

\section{Comparison with BTS classification}
\label{sec:comp}

In order to test the classification accuracy determined in this work, we need true (or accurate) classification.
As we described in Sec.~\ref{sec:intro}, BTS provides an accurate spectroscopic classification of transients based on human visual inspection and spectra from various larger telescopes.
Therefore, for the analysis below, we assume that the BTS classification is correct.

\subsection{Matching with the BTS classification}
\label{subsec:match_unmatch}

\begin{table}
\centering
\caption{Summary of the match/non-match between our classification and the BTS classification in each category.}
\label{tab:match_unmatch}
\begin{tabular}{l c c c}
\hline\hline\\[-0.8em]
                             & 2 Categories        && 5 Categories  \\  [0.2em] \cline{2-2} \cline{4-4}
                             & Match : Non-match   && Match : Non-match \\
\hline\\[-0.8em]
\snid{}                   & 3599 (79\%) : 939 (21\%)           && 2875 (63\%) :1663 (37\%)             \\ [0.30em]
\ngsf{}                   & 4068 (88\%) : 578 (12\%)           && 3476 (75\%) : 1170 (25\%)            \\ [0.30em]
\dash{}                  & 3541 (76\%) : 1105 (24\%)           && 2872 (62\%) : 1774 (38\%)            \\ [0.30em]
\hline
\end{tabular}
\end{table}

We first investigate whether our classification is matched with the BTS classification.
Tab.~\ref{tab:match_unmatch} presents a summary of the overall match/non-match between our classifications and the BTS classifications in  2 or 5 categories (see Tab.~\ref{tab:types}).
The result for 2 categories shows, on average, a $\sim$14\% better match rate than the result of 5 categories.
Among the three spectral classification tools, \ngsf{} shows the best match rate in both categories.
\snid{} and \dash{} have similar match and non-match rates.

\begin{figure*}
 \centering
  	\includegraphics[width=0.45\textwidth]{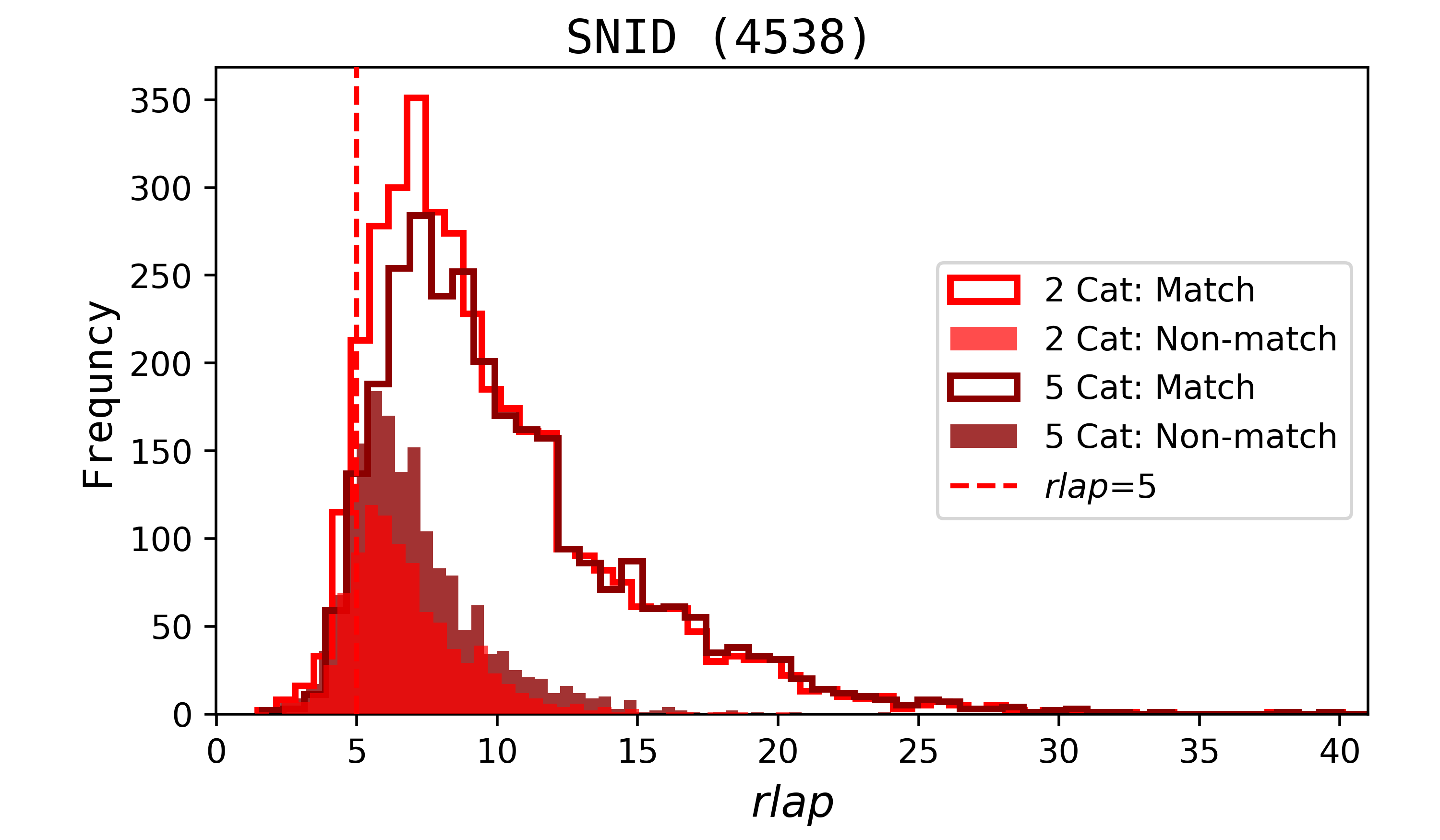}
  	\includegraphics[width=0.45\textwidth]{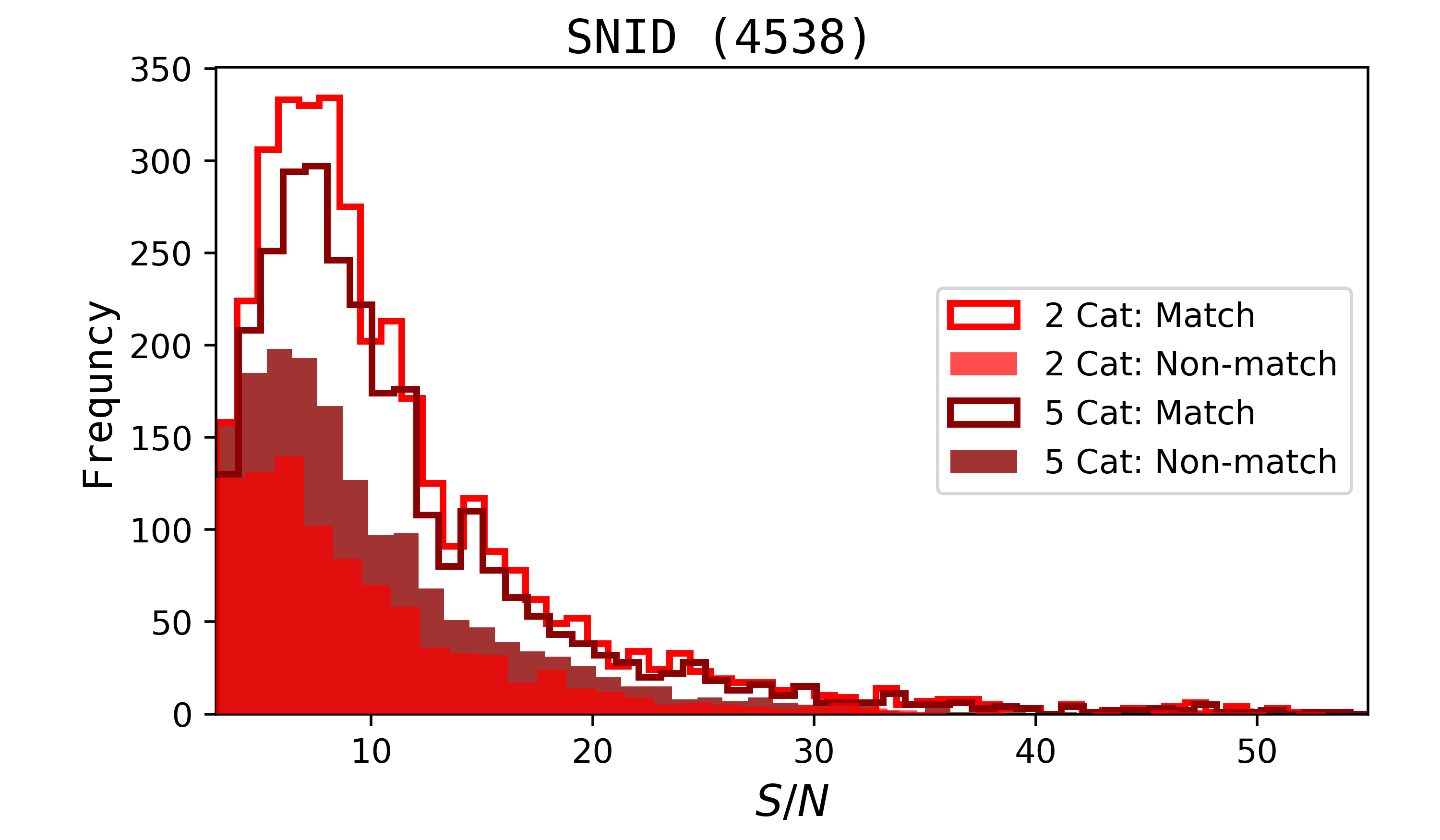}
	\includegraphics[width=0.45\textwidth]{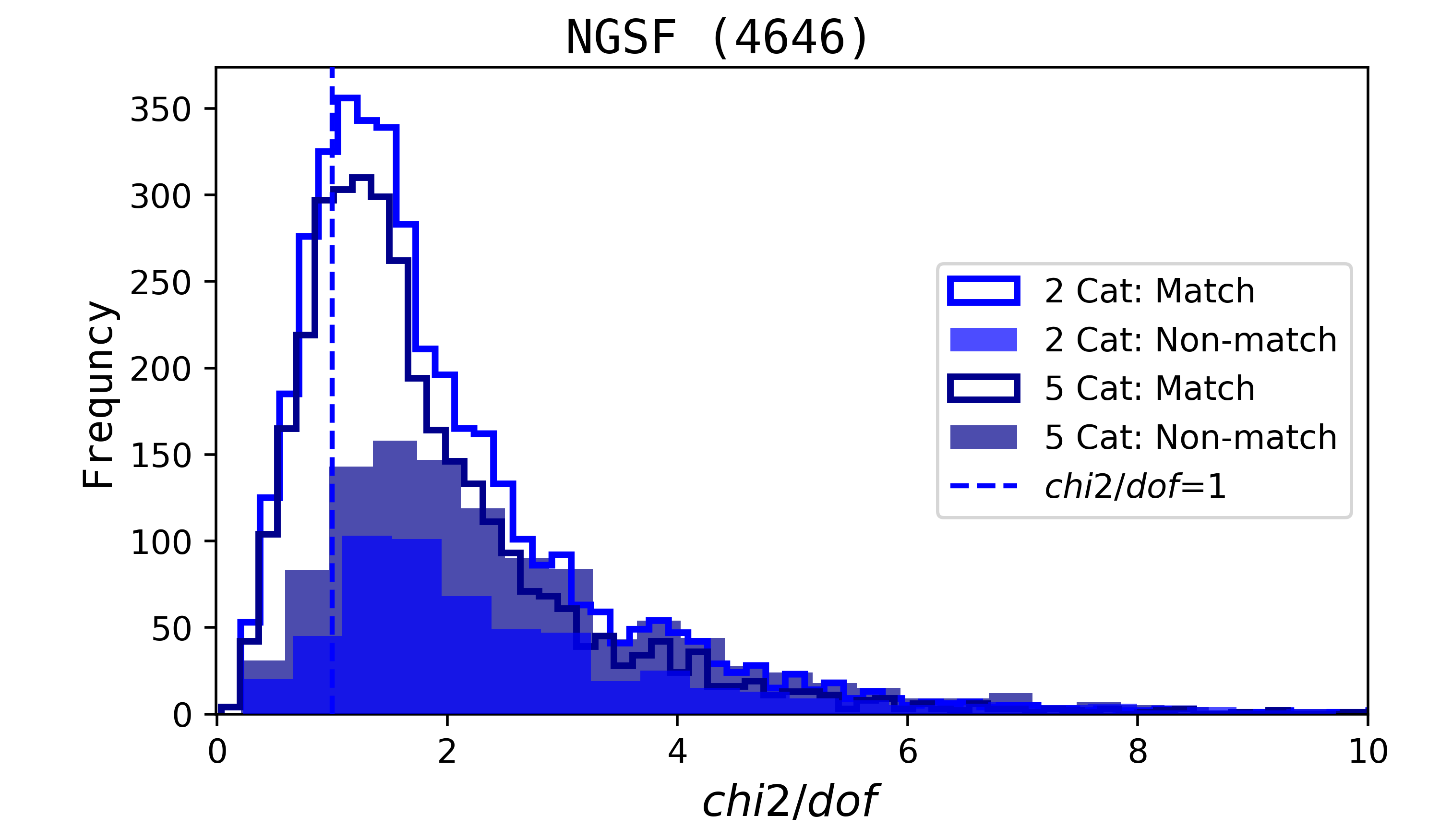}
	\includegraphics[width=0.45\textwidth]{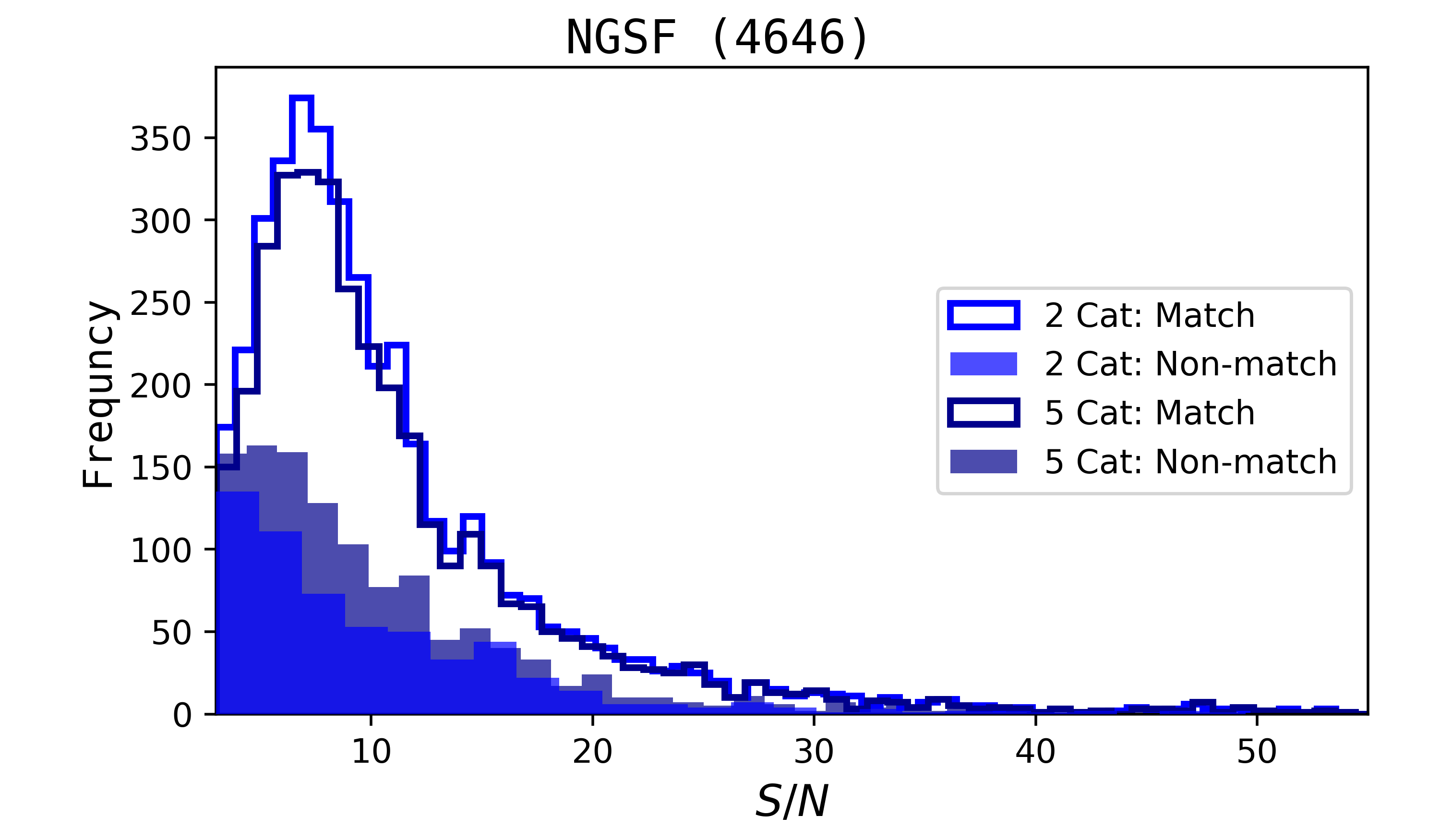}
  	\includegraphics[width=0.45\textwidth]{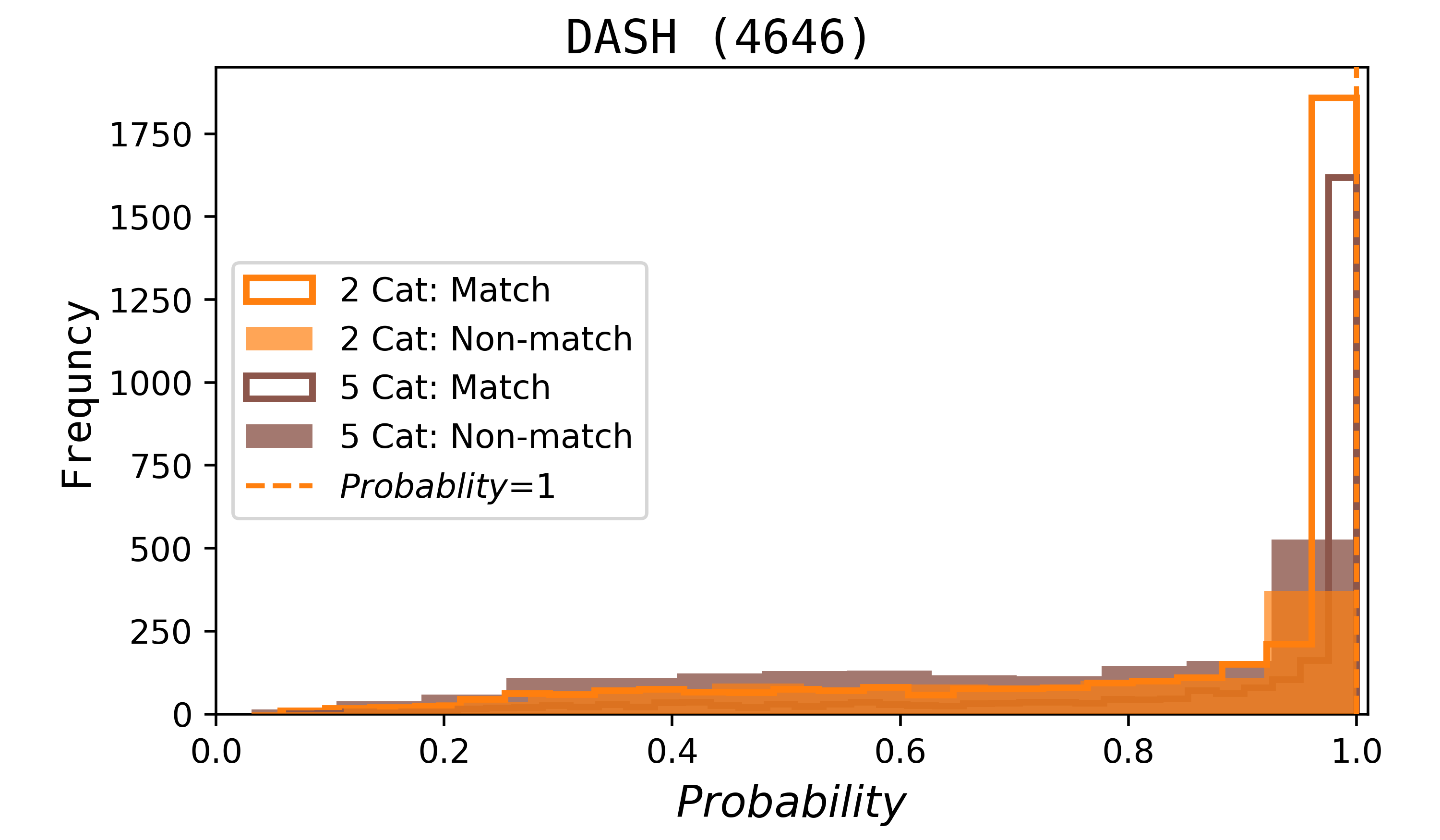}
  	\includegraphics[width=0.45\textwidth]{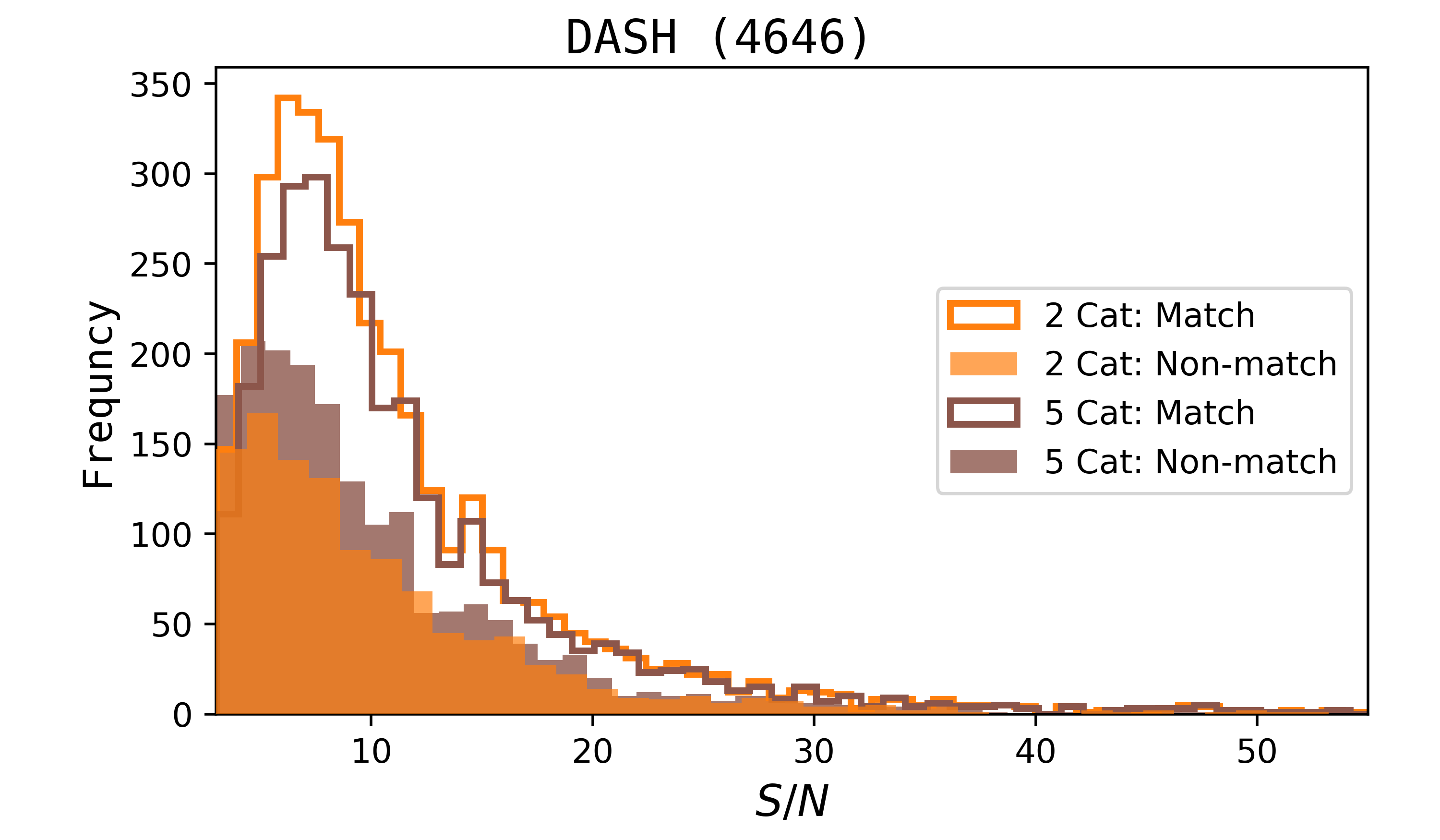}
  \caption{Distributions of match/non-match between our classification and the BTS classification as functions of each tool's primary metric and SEDM \snr{}.	
		Darker-coloured  histograms represent spectra in 5 categories, while normal colours are for 2 categories.
		Dashed lines in the left panels are the classification quality criteria: \snid{} \rrlap{} = 5 (red), \ngsf{} \chidof{} = 1.0 (blue), and \dash{} \prob{} = 1.0 (orange).
		The primary metric of each tool appears to be a more important factor than the \snr{} of spectra (i.e., quality of spectra) when it comes to
		classification accuracy.
  		}
  \label{fig:match_w_bts}
\end{figure*}

Then, we examine the distribution of match/non-match between our classifications and the BTS classifications as functions of each tool's primary metric and the \snr{} of the SEDM spectrum  in Fig.~\ref{fig:match_w_bts}.
As \ngsf{} and \dash{} metrics approach a value of 1.0 or the \snid{} \rrlap{} value has a greater value, the fraction of the matched spectra increases.
Notably, for \snid{}, our classifications match very well with those of BTS when \rrlap{} > 20.
We find some matched classifications from \ngsf{}, even though their value of \chidof{} is very high.
This may be because, by chance, the best-fitted classification with a high \chidof{} value matches the BTS classification, even though there are no well-fitted templates.
Similarly, we see in the \dash{} result that many matched classifications have low \prob{} values.
As \dash{} \prob{} is the relative probability, this value may seem low if \dash{} has several similar classifications and it spreads its \prob{} evenly across them. 

With \snr{}, the three classifiers show similar distributions.
The shape of the distribution follows that of \snr{} (see Fig.~\ref{fig:sample_distribution}).
There are some unmatched spectra with high \snr{} values.

No difference in each tool's primary metric and \snr{} distributions is observed between the distributions of 2 and 5 categories.

The results of this section suggest that the primary metric of each tool appears to be a more important factor than the \snr{} of spectra (i.e., quality of spectra) when it comes to classification accuracy.

\subsection{Overall confusion matrix and performance evaluation metrics}
\label{subsec:overall_cf}

\begin{figure*}
 \centering
  	\includegraphics[width=0.49\textwidth]{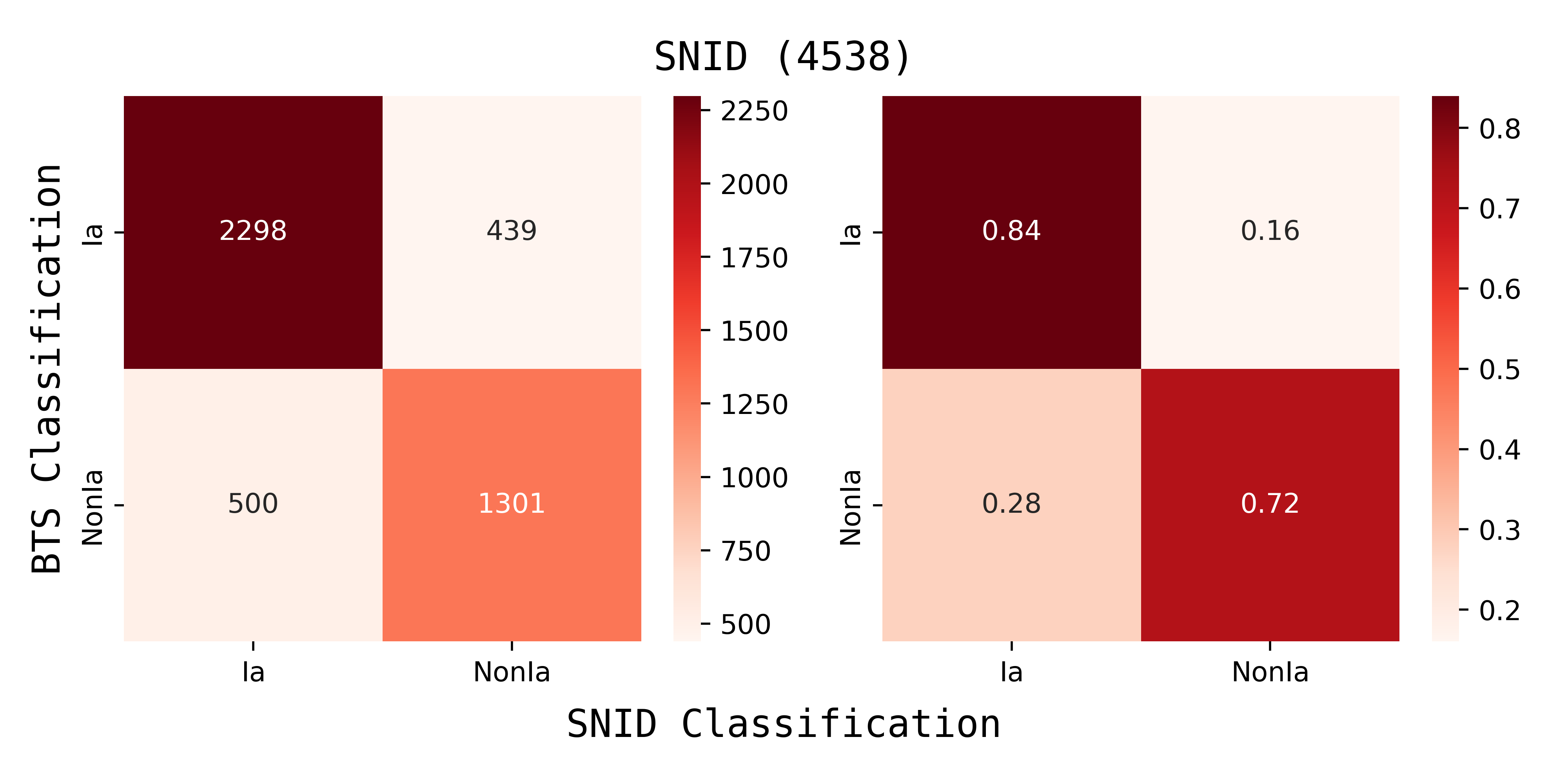}
  	\includegraphics[width=0.49\textwidth]{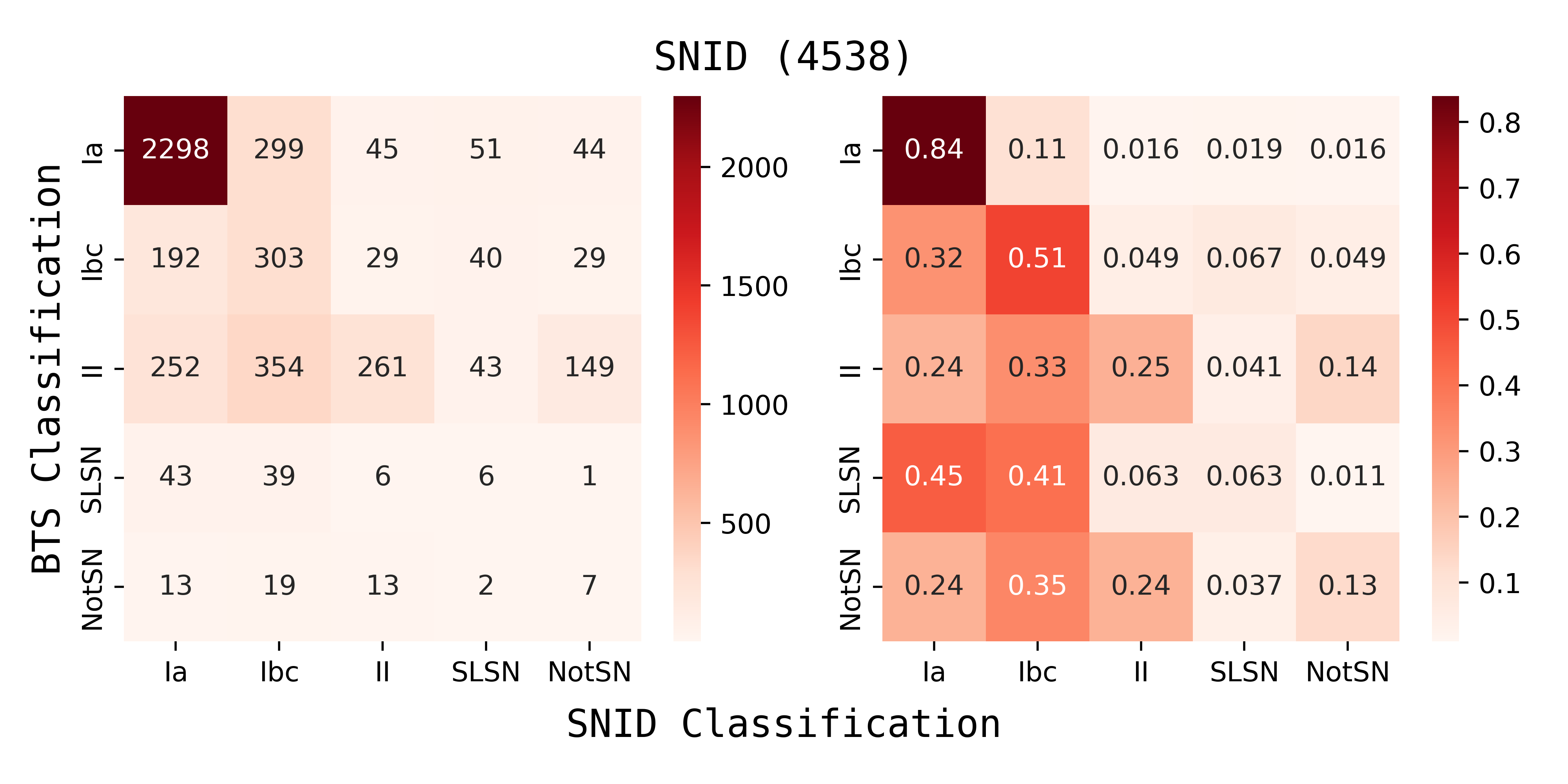}
	\includegraphics[width=0.49\textwidth]{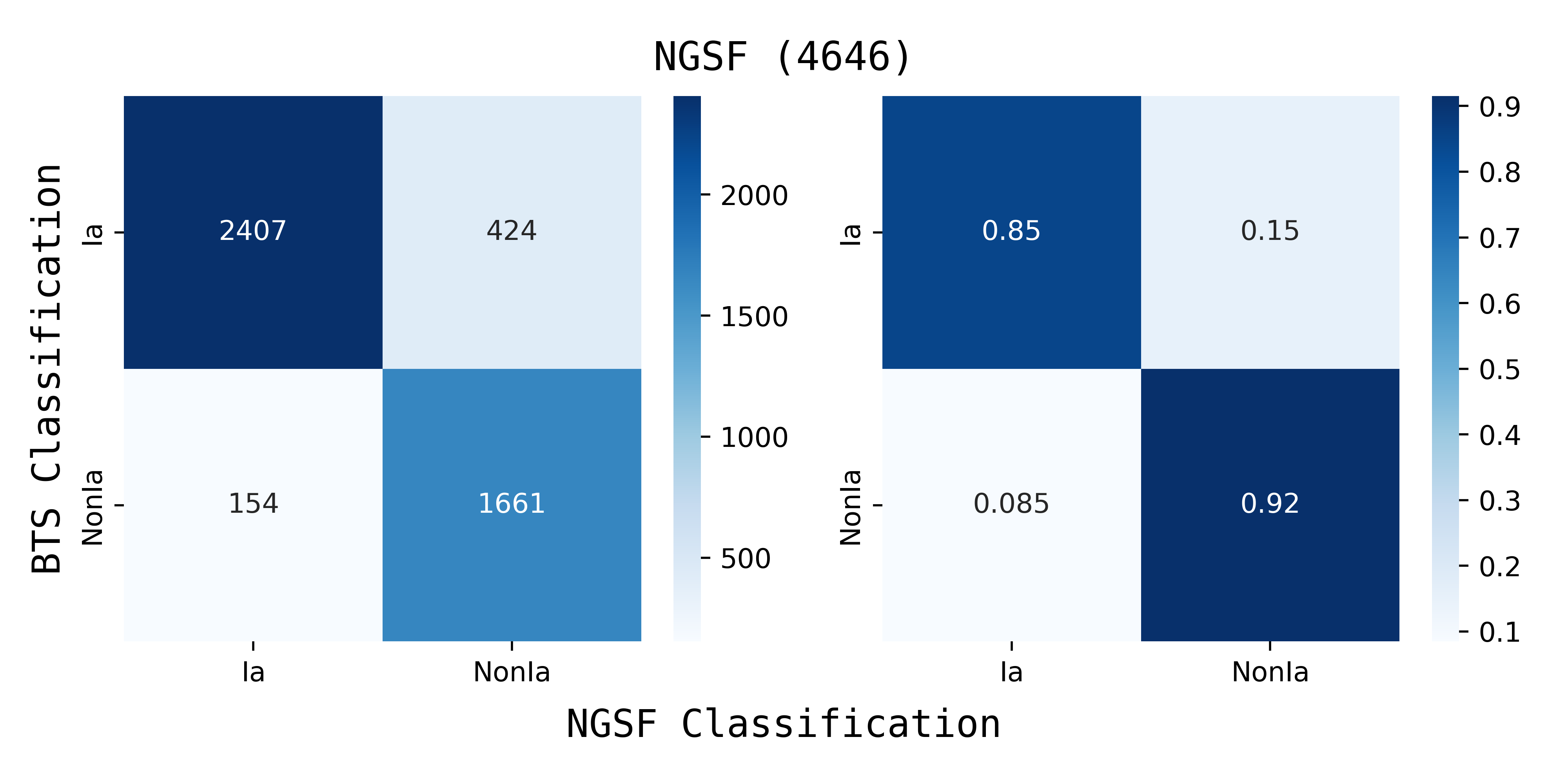}
	\includegraphics[width=0.49\textwidth]{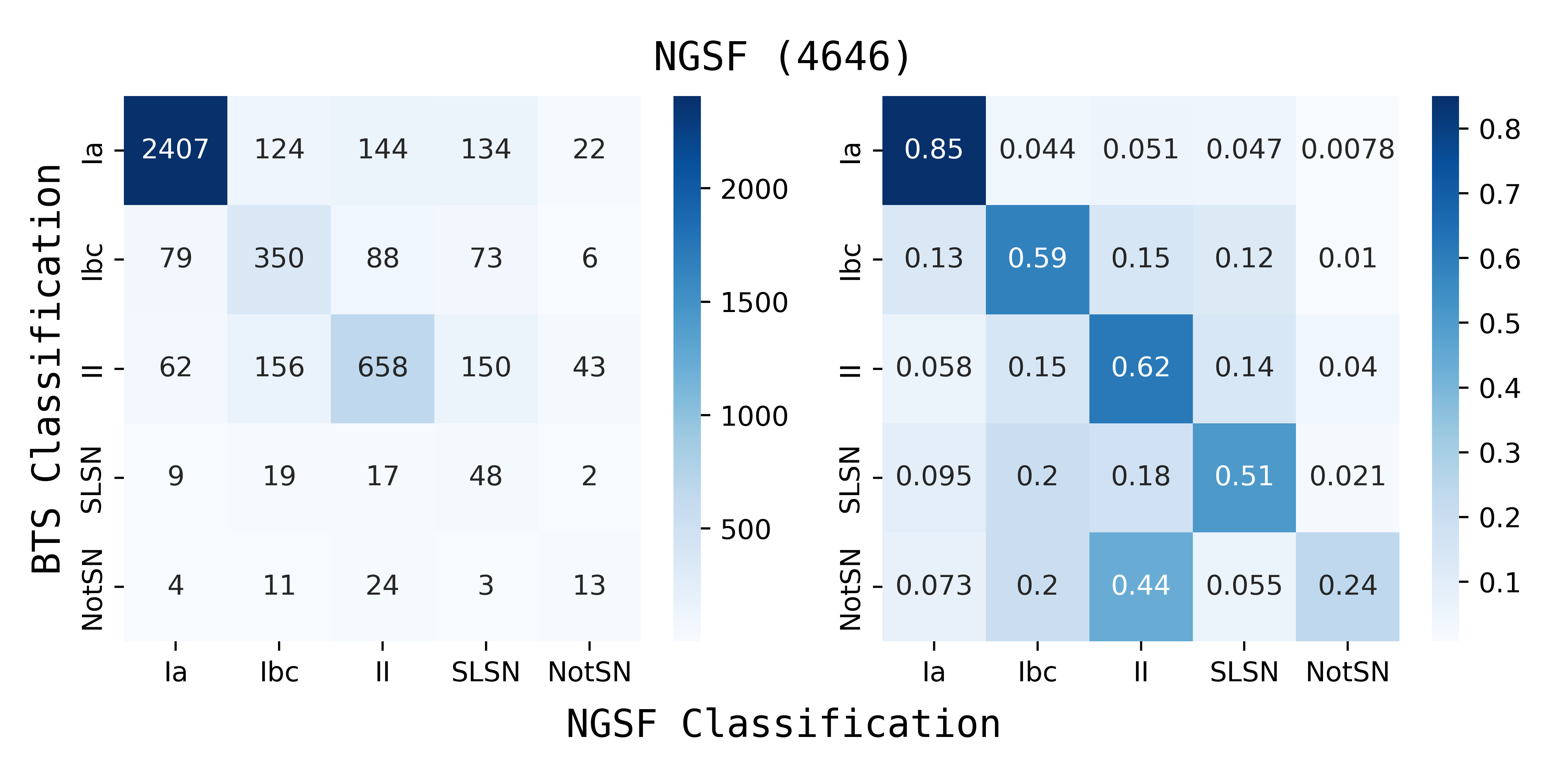}
  	\includegraphics[width=0.49\textwidth]{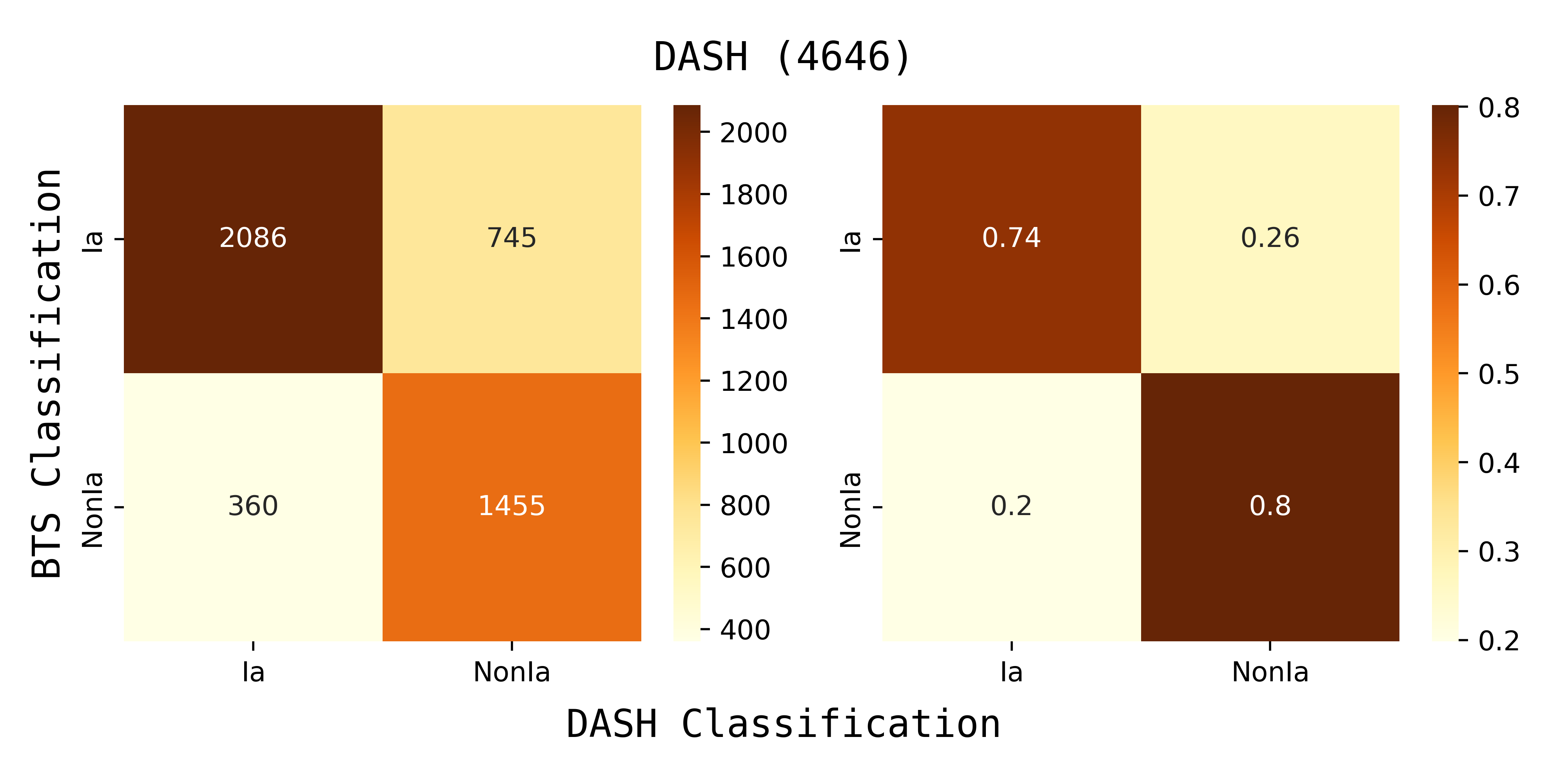}
  	\includegraphics[width=0.49\textwidth]{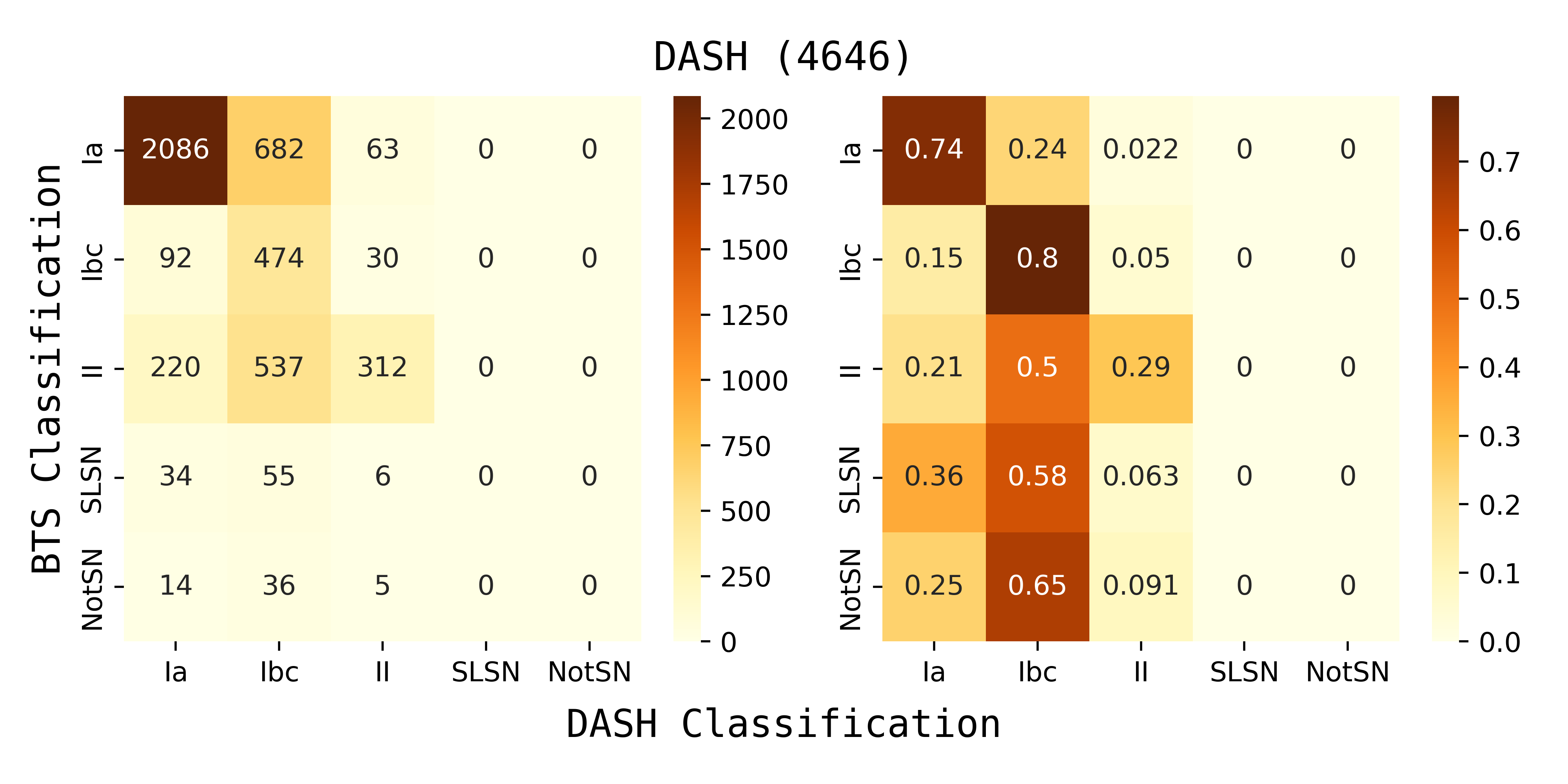}
  \caption{Confusion matrices for each classification tool split by 2 (left panel) and 5 (right) categories. 
  		In each panel, the left plot shows an absolute number of each classification, while the right plot presents the normalized confusion matrix, i.e. the fraction of each BTS classification (the True label; y-axis) that was classified from this work (the Predicted label; x-axis).
		This normalized confusion matrix is normalized to 1 along rows.
		The diagonal elements indicate correctly classified spectra, while off-diagonal elements are the spectra that are misclassified by this work.
		\ngsf{} has the best performance, while \snid{} and \dash{} have similar performance.
		We note that \dash{} does not have the option to classify an object as SNLS or NotSN types. 
  		}
  \label{fig:overall_cf}
\end{figure*}

\begin{table*}
\centering
\caption{Overall performance evaluation metrics for each classification tool split by type in each category.}
\label{tab:eval_metrics}
\begin{tabular}{l c c c c c c c c c}
\hline\hline\\[-0.8em]
                                  &                                  & \multicolumn{2}{c}{2 Categories}  && \multicolumn{5}{c}{5 Categories} \\  [0.2em] \cline{3-4} \cline{6-10}
                                  &                                  & Ia                   & NonIa                   && Ia           & Ibc           & II         & SLSN  & NotSN \\
\hline\\[-0.8em]
\multirow{4}{*}\snid{}  & \textit{Accuracy}                  &  \multicolumn{2}{c}{79.3\%}           &&  \multicolumn{5}{c}{63.4\%}           \\ [0.30em]
                                  & \textit{Purity}                        & 82.1\%           & 74.8\%                  && 82.1\%  & 29.9\%   & 73.7\%  & 4.2\%  & 3.0\%   \\ [0.30em]
                                  & \textit{TPR} 			   & 84.0\%	    & 72.2\%		   &&  84.0\%  & 51.1\%  & 24.6\%  & 6.3\%  & 13.0\% \\ [0.30em]
				& \textit{FPR}			   &  27.8\%  	    & 16.0\%		   &&  27.8\%  & 18.0\%    & 12.7\%  & 3.1\%  & 5.0\%  \\ [0.30em]
\hline
\multirow{4}{*}\ngsf{}  & \textit{Accuracy}                  &  \multicolumn{2}{c}{87.6\%}           &&  \multicolumn{5}{c}{74.8\%}           \\ [0.30em]
				 & \textit{Purity}                        & 94.0\%           & 79.7\%                  && 94.0\%  & 53.0\%   & 70.7\%  & 11.8\%  & 15.1\%   \\ [0.30em]
				 & \textit{TPR} 			   & 85.0\%	    & 91.5\%		   &&  85.0\%  & 58.7\%  & 61.6\%  & 50.5\%  & 23.6\% \\ [0.30em]
				 & \textit{FPR}			   &  8.5\%  	    & 15.0\%		   &&  8.5\%   & 7.7\%     & 7.6\%    & 7.9\%    & 1.6\%  \\ [0.30em]
\hline
\multirow{4}{*}\dash{} & \textit{Accuracy}                  &  \multicolumn{2}{c}{76.2\%}           &&  \multicolumn{5}{c}{61.8\%}           \\ [0.30em]
				 & \textit{Purity}                        & 85.3\%           & 66.1\%                  && 85.3\%  & 26.6\%   & 75.0\%  & -            & -   \\ [0.30em]
				 & \textit{TPR} 			    & 73.7\%	    & 80.2\%		   &&  73.7\%  & 79.5\%  & 29.2\%  & -            & -   \\ [0.30em]
				 & \textit{FPR}			   &  19.8\%  	    & 26.3\%		   &&  19.8\%  & 32.3\%    & 2.9\%  & -            & -   \\ [0.30em]
\hline
\end{tabular}
\end{table*}

For a more detailed analysis of classification performance, we illustrate the confusion matrices in Fig.~\ref{fig:overall_cf} and then derive evaluation metrics in Tab~\ref{tab:eval_metrics}.
For the performance evaluation metrics, we use `\textit{Accuracy}', `\textit{Purity}', `\textit{True Positive Rate (TPR)}', and `\textit{False Positive Rate (FPR)}'.
\textit{Accuracy} is calculated as the number of correctly classified spectra against the total number of spectra.
\textit{Purity} is the fraction of correctly classified spectra among all classified spectra in the same type.
\textit{TPR}, also called classification efficiency, shows the probability of true classification, while \textit{FPR} presents the probability of false classification in each type.
Each evaluation metric is calculated as follows:

\begin{equation}
\label{eq:accuracy_2cat}
\textit{Accuracy = } \frac{\textit{TP + TN}} {\textit{TP + TN + FP + FN}} \text{\; for 2 categories;}
\end{equation}

\begin{equation}
\label{eq:accuracy_5cat}
\textit{\qquad \qquad = } \frac{\text{Total number of true positive}} {\text{Total number of spectra}}  \text{\; for 5 categories;}
\end{equation}

\begin{equation}
\label{eq:purity}
\textit{Purity = } \frac{\textit{TP}} {\textit{TP + FP}} \text{;}
\end{equation}

\begin{equation}
\label{eq:tpr}
\textit{TPR = } \frac{\textit{TP}} {\textit{TP + FN}} \text{;}
\end{equation}

\begin{equation}
\label{eq:fpr}
\textit{FPR = } \frac{\textit{FP}} {\textit{FP + TN}} \text{;}
\end{equation}

where TP and TN  are the numbers of true positives and true negatives, respectively, and FP and FN are false positive and false negative, respectively.
In the analysis below, we calculate a separate \textit{Purity}, \textit{TPR}, and \textit{FPR} for each classification.

When the predicted classification from this work is consistent with the true BTS classification, the diagonal elements in the confusion matrix, which indicate correctly classified in this work, will be coloured darker, while the off-diagonal elements, which indicate misclassification, will be coloured close to white in the normalized confusion matrix (right plots in each panel).
In this respect, at first glance, the normalized confusion matrices of \ngsf{} follow this diagonal trend, while those of \snid{} and \dash{} do not.
When we consider the confusion matrices of the 2 categories, they also seem to follow the diagonal trend.
However, those of the 5 categories appear scattered, except for the \ngsf{} one.

Broadly speaking, in the results of the 2 categories, all tools show similar \textit{Accuracy} ($\sim$81\%) in the classification of SNe Ia and NonIa.
However, considering the \textit{Purity}, which is strongly correlated to \textit{TPR} and anti-correlated to \textit{FPR}, all tools show better results for SN Ia classification.
Notably, \ngsf{} shows the best result with \textit{Purity} = 94\% and \textit{FPR} = 8.5\%.

Regarding the results of the 5 categories, \ngsf{} shows the best performance when evaluating based on the confusion matrix and performance evaluation metrics.
The well-distributed dark colours in \snid{} and \dash{} normalized confusion matrices indicate that \snid{} and \dash{} classify most of the spectra into SNe Ia (62\% and 53\%, respectively) and Ibc (22\% and 38\%) types.
In total, 84\% of \snid{} and 91\% of \dash{} results are classified as SNe Ia and Ibc (cf. 69\% for \ngsf{}), compared to 72\% for the true BTS types.
This shows the so-called ``type attractor'' issue (see \citealt{Blondin2007, Muthukrishna2019}; \citetalias{Kim2022}).
The imbalance of templates biases the classification toward a type that has more templates, when the \snr{} of spectra is low.
In the templates of \snid{} and \dash{}, $\sim$63\% are SNe Ia and $\sim$20\% are SNe Ibc, while the \ngsf{} templates have $\sim$29\% and $\sim$31\%, respectively.

From this section, we find that \ngsf{} shows the best performance among the three spectral classification tools when evaluating based on the confusion matrix and performance evaluation metrics.
\snid{} and \dash{} have similar performance, even though \dash{} has no SLSN and NotSN types.
In addition, we can say that splitting a sample into 2 categories gives a more accurate classification than when splitting a sample into 5 categories.
This is because classifying transients other than SNe Ia is more uncertain than classifying SNe Ia from spectral classification tools alone.

\subsection{Performance evaluation metrics split by each tool's primary metric}
\label{subsec:eval_metrics}

\begin{table}
\centering
\caption{The sample size split by each tool's primary metric.}
\label{tab:grouping}
\begin{tabular}{c c c c c c c c}
\hline\hline\\[-0.8em]
\multicolumn{2}{c}{\snid}        && \multicolumn{2}{c}{\ngsf}                    && \multicolumn{2}{c}{\dash}     \\[0.2em] \cline{1-2} \cline{4-5} \cline{7-8}
\rrlap{}          &  $N$                && \chidof{}        & $N$                            && \prob{}        & $N$           \\[0.15em] 
\hline\\[-0.8em]
[0, 4)             &  81                  &&  [0.00, 0.50)  & 152                            && [0.00, 0.30)        & 268           \\ [0.30em]
[4, 5)             &  304                &&  [0.50, 0.75)  & 298                            && [0.30, 0.40)        & 249           \\ [0.30em]
[5, 6)             &  613        	     &&  [0.75, 1.00)  & 471                             && [0.40, 0.50)        & 290           \\ [0.30em]
[6, 7)             &  676               &&  [1.00, 1.25)  & 582                             && [0.50, 0.60)        & 275           \\ [0.30em]
[7, 8)             &  590               &&  [1.25, 1.50)  & 561                             && [0.60, 0.70)        & 282           \\ [0.30em]
[8, 9)             &  486               &&  [1.50, 1.75)  & 501                             && [0.70, 0.80)        & 308           \\ [0.30em]
[9, 10)           &  366               &&  [1.75, 2.00)  & 339                             && [0.80, 0.90)        & 413           \\ [0.30em]
[11, 12)         &  268               &&  [2.00, 2.25)  & 295                             && [0.90, 0.92)        & 114           \\ [0.30em]
[12, 13)          &  160              &&  [2.25, 2.50)  & 258                             && [0.92, 0.94)        & 135           \\ [0.30em]
[13, 14)          &  127              &&  [2.50, 3.00)  & 353                             && [0.94, 0.96)        & 151           \\ [0.30em]
[14, 15)          &  119              &&  [3.00, 3.50)  & 212                             && [0.96, 0.98)        & 252           \\ [0.30em]
[15, 16)          &  88                &&  [3.50, 4.00)  & 174                             && [0.98, 0.99)        & 193           \\ [0.30em]
[16, 17)          &  88               &&  [4.00, 4.50)  & 118                              && [0.99, 1.00)        & 1058           \\ [0.30em]
[17, 18)          &  59               &&  [4.50, 5.00)  & 88                                && [1.00]                 & 658           \\ [0.30em]
[18, 19)          &  50               &&  [5.00, inf)           & 244      \\ [0.30em]
[19, 20)          &  49        \\ [0.30em]
[20, inf)               &  125        \\ [0.30em]
\hline
\end{tabular}
\end{table}

\begin{figure*}
 \centering
  	\includegraphics[width=0.49\textwidth]{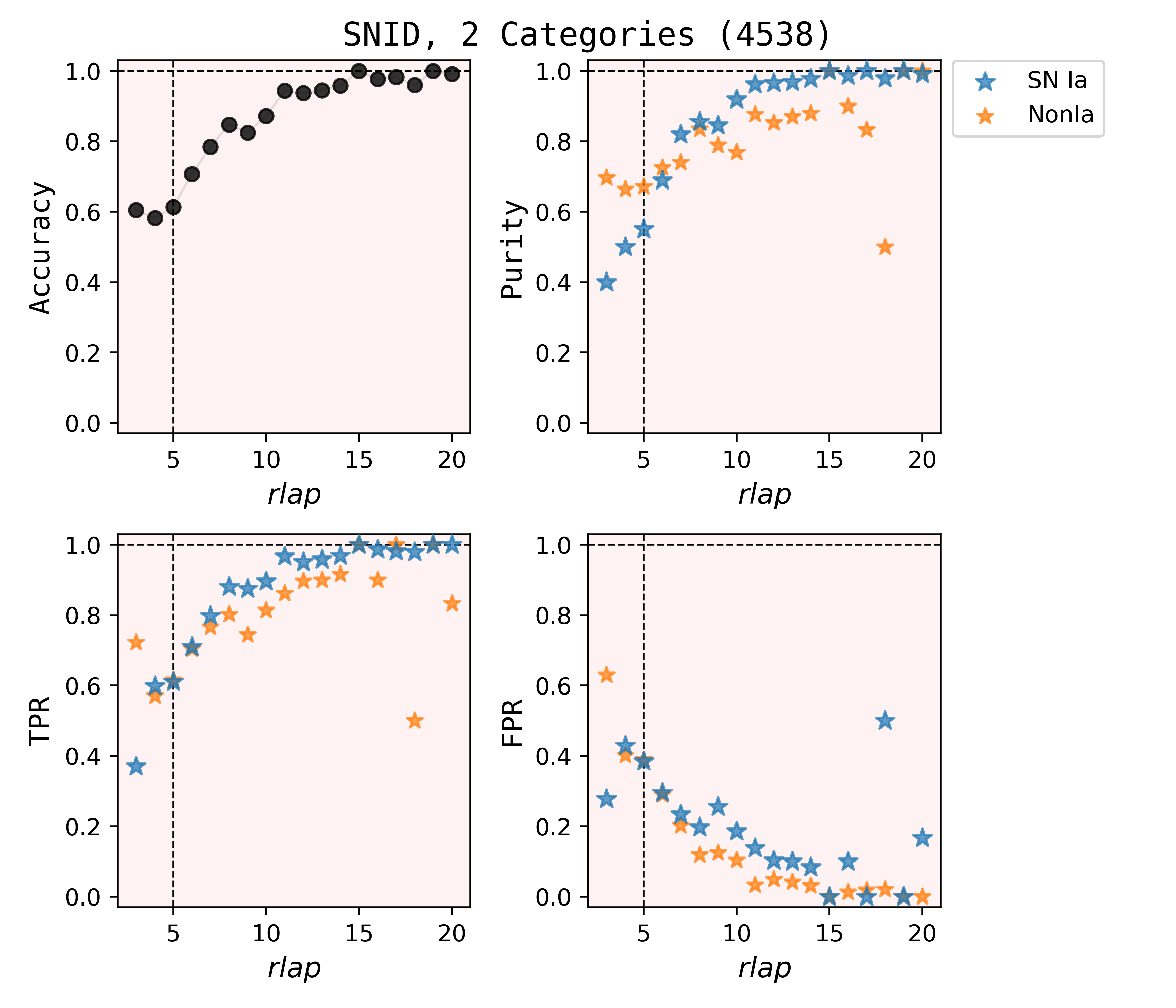}
  	\includegraphics[width=0.49\textwidth]{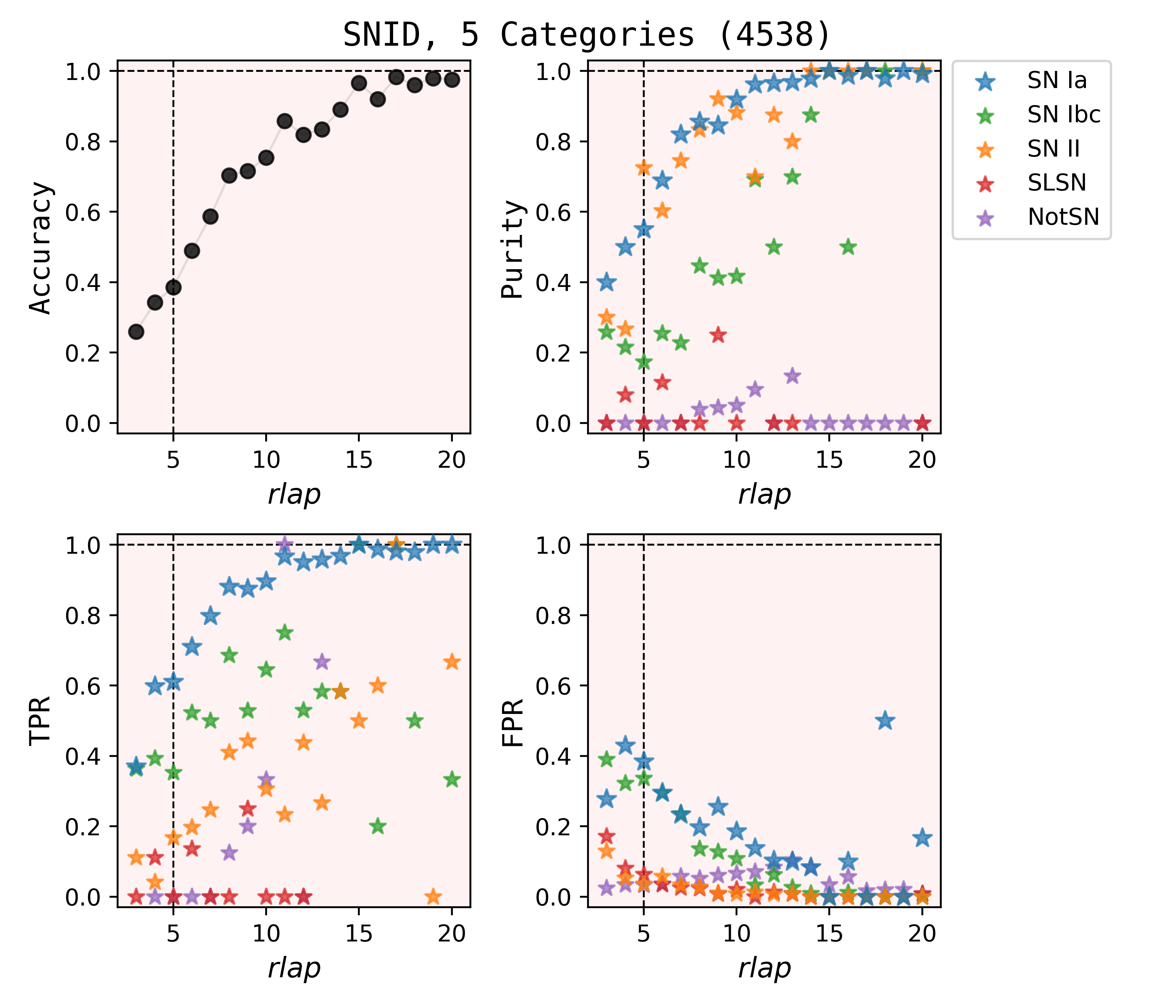}
	\includegraphics[width=0.49\textwidth]{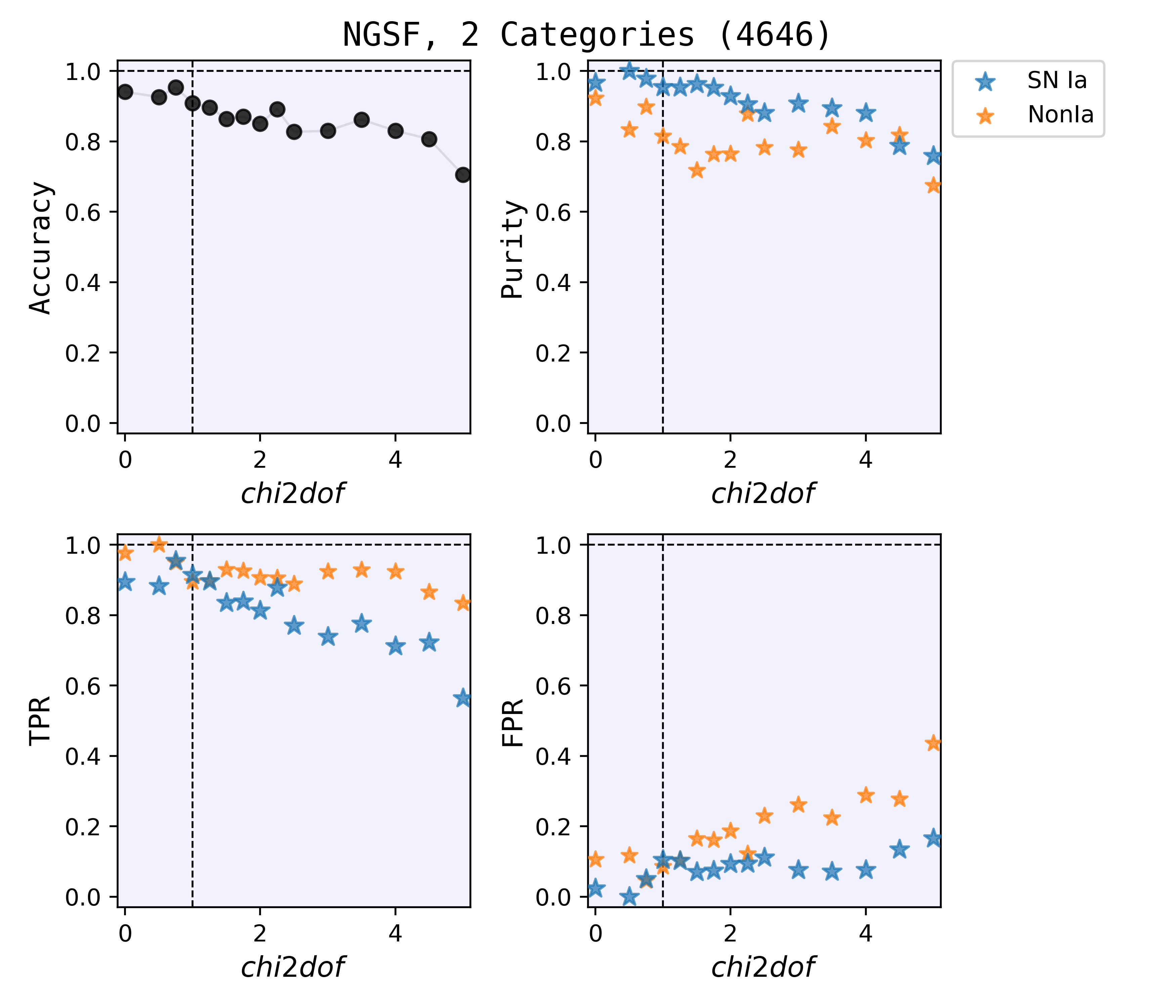}
	\includegraphics[width=0.49\textwidth]{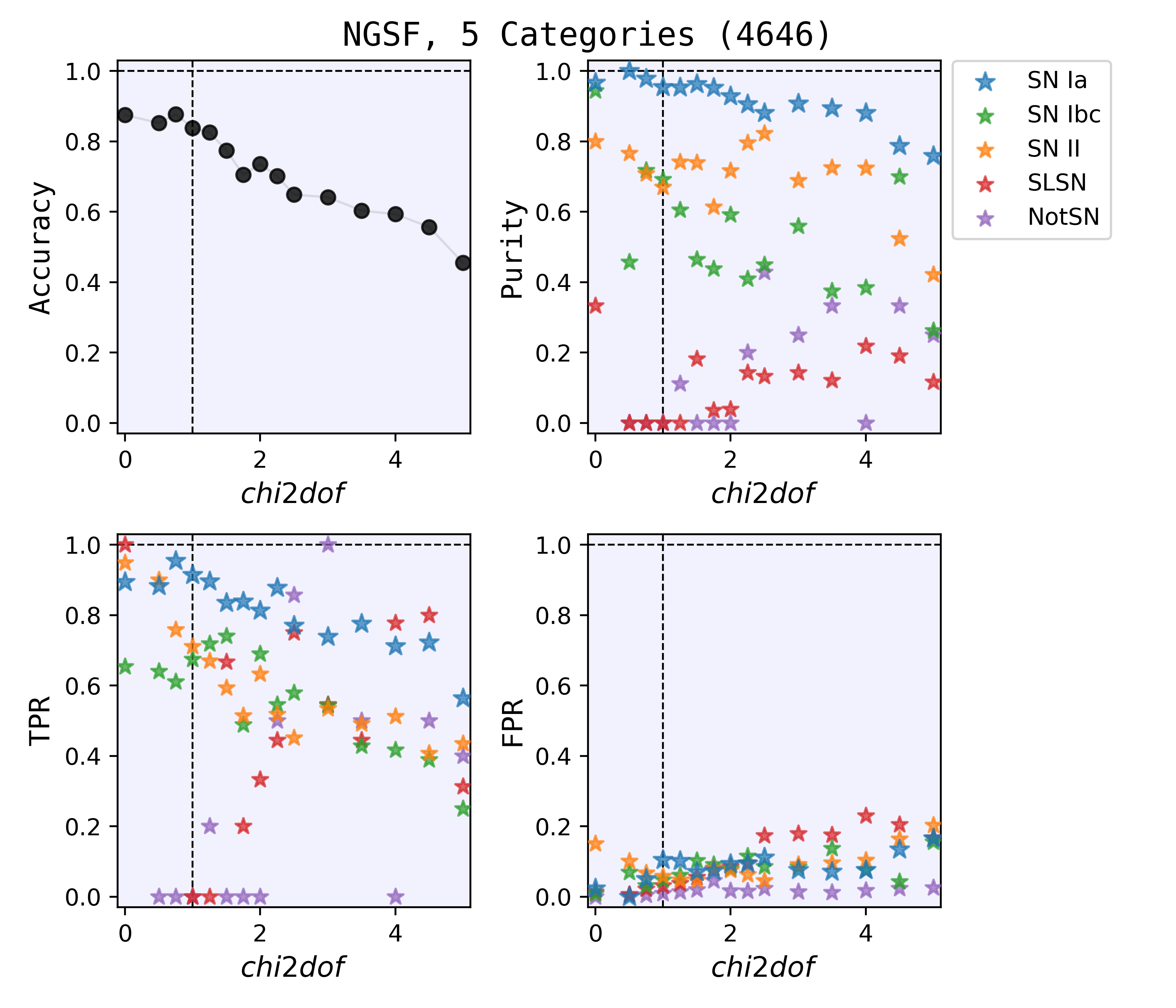}
  	\includegraphics[width=0.49\textwidth]{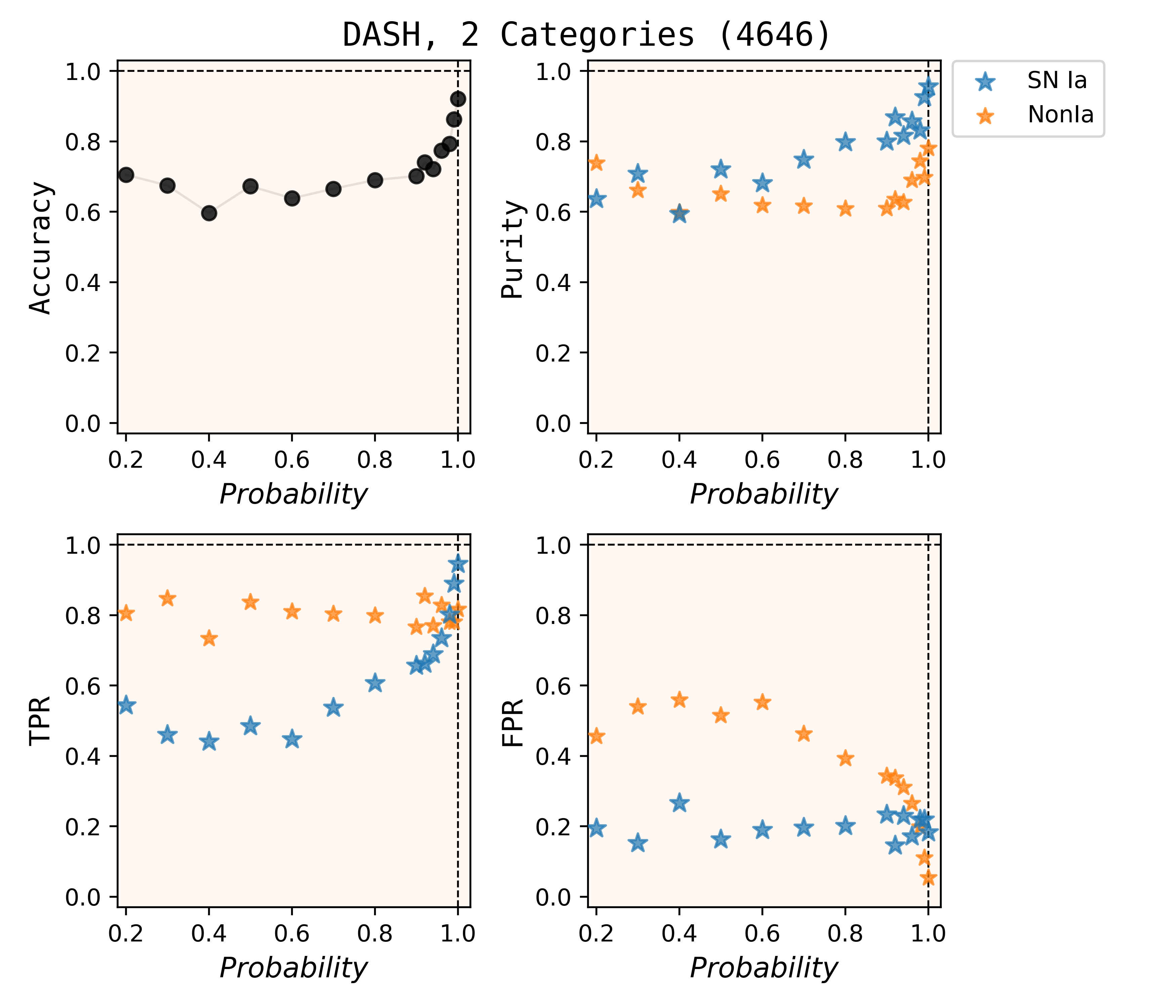}
  	\includegraphics[width=0.49\textwidth]{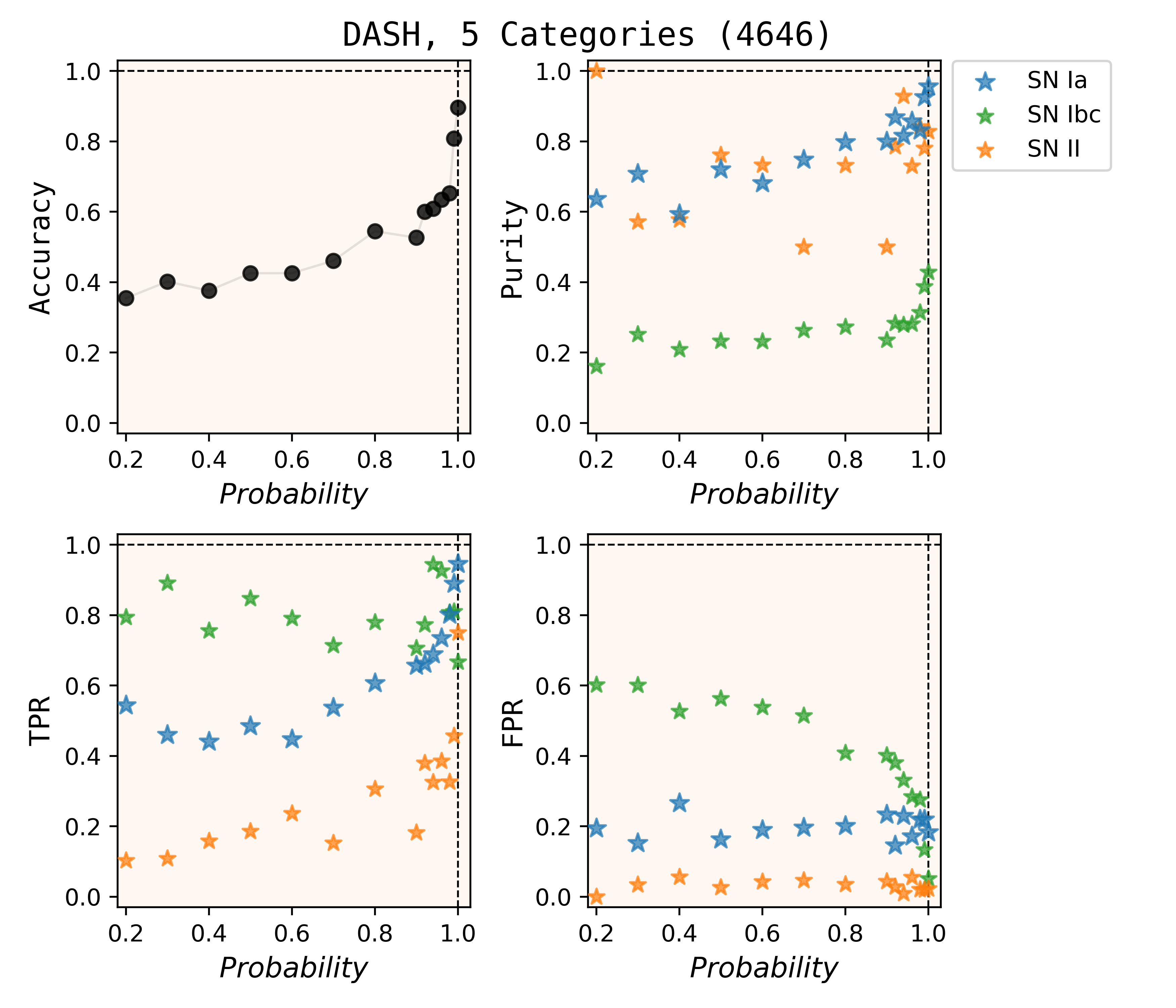}
  \caption{Performance evaluation metrics for each tool split by its primary metric: \snid{} by \rrlap{} (top panel), \ngsf{} by \chidof{} (middle), and \dash{} by \prob{} (bottom).
  		}
  \label{fig:eval_metrics}
\end{figure*}

Additionally, we examine the accuracy of each tool's primary metric.
For this, we split the sample by each tool's primary metric (Tab.~\ref{tab:grouping}) and separately calculate performance evaluation metrics for the sample in each bin.
When we split the sample, we use smaller step sizes for denser bins, e.g., $\Delta$\chidof{} = 0.25 for \ngsf{} \chidof{} between 0.50 and 2.50, and $\Delta$\prob{} = 0.01 or 0.02 for \dash{} \prob{} between 0.90 and 1.00.
We note that the sample size in some types, particularly the NotSN type in 5 categories, is small.

Fig.~\ref{fig:eval_metrics} shows this calculation: performance evaluation metrics for each tool split by its primary metric.

\begin{description}
\item[-\snid{}:]
Approaching the higher value of \rrlap{} improves all the evaluation metrics.
The most populated range of \rrlap{} is between 5 and 10, with a median of 8.0 (see the top panel of Fig.~\ref{fig:quality_value_distribution}).
In this region, \textit{Accuracy}, \textit{Purity}, and \textit{TPR} are between 60\% and 80\%, while \textit{FPR} is between 20\% and 40\% for 2 categories.
The evaluation metrics of the 5 categories are worse.
Notably, when \rrlap{} > 15, \snid{} can accurately classify SNe Ia without contamination from other types (\textit{Purity} > 98\%), similar to the result of Sec.~\ref{subsec:match_unmatch}.
SNe Ibc and II are also classified accurately, but it is most likely to be driven by the small number of objects (< 10) when \rrlap{} > 15.
Classifying SLSNe seems to be very inaccurate with \snid{}.
\end{description}

\begin{description}
\item[-\ngsf{}:]
As \chidof{} approaches 1.0, all evaluation metrics show improvement with a peak at the [0.75, 1.00) bin.
\chidof{} between 0.5 and 2, the most populated region (see the middle panel of Fig.~\ref{fig:quality_value_distribution}), gives a reasonable accuracy of classification: \textit{Accuracy} of $\sim$90\% for 2 categories and $\sim$80\% for 5 categories.
Especially, \textit{FPR} for all types in both categories has a low value of $\sim$10\%.
\ngsf{} can classify SNe Ia well when considering all the metrics, while that is not the case for NonIa types.
SLSNe cannot be classified by \ngsf{}.
\end{description}

\begin{description}
\item[-\dash{}:]
As \prob{} approaches 1.0, all the evaluation values show improvement.
Notably, \textit{Purity} of SNe Ibc is especially low compared to other types (see also Tab.~\ref{tab:eval_metrics}).
This is because \dash{} has classified a lot of spectra into the SN Ibc type, especially for the true BTS SLSN and NotSN types, for which there are no templates in the current version of \dash{} (see the rightmost corner plot of Fig.~\ref{fig:overall_cf}).
\end{description}

\section{Application: supporting metrics for spectroscopic classification of transients}
\label{sec:app}

The performance test results in this work show that it is difficult to obtain an accurate classification from spectral classification tools alone.
Thus, for example, the BTS team, which used \snid{} as a preliminary classification tool, used visual inspection to securely classify the spectra manually.
However, this approach is very time-consuming when classifying thousands of transients and requires human resources to do it on a daily basis.

In an attempt to address this, the BTS team have developed and implemented \sniascore{}, a deep-leaning-based tool optimized for classifying only SNe Ia from SEDM spectra, to reduce human visual inspection by $\sim$60\% \citep[see ][for a discussion in detail]{Fremling2021}.
In the current situation, we expect that our work can also contribute to reducing human visual inspection by providing supporting metrics for future spectroscopic follow-up surveys.

When a spectrum of a new transient is classified by one of the spectral classification tools, the tool will return a type with its primary metric.
For example, if \snid{} gives `Ia' with \rrlap{} = 5.5, from our performance evaluation metrics for 2 Categories, we can evaluate the accuracy of the \snid{} classification, to be \textit{Accuracy} = 61\%, \textit{Purity} = 55\%, \textit{TPR} = 61\%, and \textit{FPR} = 38\%.
By combining these \snid{} metrics with other tools' metrics, we can determine which targets require visual inspection.

Our suggestion is that for real observations, a high \textit{Purity} (i.e., minimising contamination) would be the most important factor when we do not know the true classification.
Since \textit{Purity} is strongly correlated to \textit{TPR} and anti-correlated to \textit{FPR}, \textit{Purity} can provide a balanced measure between them.
Therefore, for example, the determined type is most likely accurate when all three tools give the same transient type and \textit{Purity} > 90\%.
Another suggestion is low enough \textit{FPR} (<1\%) so that the classification determined by tools does not require human visual inspection, as discussed by \citet{Fremling2021}.

\section{Discussion}
\label{sec:discussion}

In this work, we have examined the accuracy of classification for widely used spectral classification tools, such as \snid{}, \ngsf{} (a Python version of \superfit{}), and \dash{}, in various ways.
For this, we used 4,646 SEDM spectra (of 2,986 individual targets) that have accurate classifications from BTS.
Comparing our classification with that of BTS, we have presented 1) the match/non-match distributions as functions of each tool's primary metric (\snid{} \rrlap{}, \ngsf{} \chidof{}, and \dash{} \prob{}) and the \snr{} of the SEDM spectrum, 2) the confusion matrices together with performance evaluation metrics, and 3) performance evaluation metrics split by each tool's primary metric for each tool.

We found that:

\begin{itemize}
	\item The primary metric of the spectral classification tool is a more important factor than \snr{} of spectra when considering the classification accuracy.
	\item Among the three tools, \ngsf{} provides the best overall classification (overall \textit{Accuracy} up to 87.6\%).
	\item \snid{} and \dash{} have similar performance with overall \textit{Accuracy} of up to 79.3\% and 76.2\%, respectively. 
	\item For SNe Ia, \snid{} can accurately classify them when \rrlap{} > 15 with \textit{Purity} >  98\%.
	\item For other types, determining their type is uncertain when using any of these spectral classification tools alone.
\end{itemize}

\snid{}, \ngsf{}, and \dash{} provide helpful information on the classification of newly discovered transients.
However, this work shows that it is difficult to obtain an accurate classification from those tools alone.
Because of this, visual inspection is needed to confirm the classification from the tools (e.g., see \citealt{Fremling2021} for BTS and \citealt{Smith2020} for the Dark Energy Survey SN Program).
To reduce this human visual inspection, this work provides supporting metrics for a classification, which are constructed based on the performance evaluation metrics split by each tool's primary metric.

One concern may arise about the impact of different spectral resolutions of the instruments when referring to our supporting metrics, which are calculated based on SEDM spectra ($R\sim100$).
SN spectral features are the key factor in classifying a transient's spectrum. 
Those features have typical widths of $\sim$100--150 $\AA$, due to the large expansion velocities of the SN ejecta \citep[$\sim$10,000 km/s;][]{Blondin2007}.
\citet{Rigault2019} showed that by comparing SEDM spectra with higher resolution spectra obtained with different instruments, the main characteristic features, such as Si II $\lambda$6500 for the SNe Ia and the P-Cygni H$\alpha$ shape for the SNe II, are clearly visible in the SEDM spectra.
In addition, the overall shape of the spectra matches well between SEDM and other higher-resolution spectra.
Furthermore, for \snid{} and \dash{}, when running the tools, the input spectrum has been preprocessed to ensure that the input spectrum has the same wavelength range with the same number of bins as the templates, making it easy to compare and classify them.
During this preprocessing, including a logarithmic wavelength binning and a band pass filtering, most of the narrower features are smoothed out \citep[see][for examples]{Blondin2007, Muthukrishna2019}.
In the case of \ngsf{}, the algorithm fits the observed spectra using binned templates (the default value is 10 $\AA$, and 20 and 30 $\AA$ binned templates are also provided), which could remove narrower features.
Therefore, we expect that the effect of resolution is not significant.
Because of the low-resolution of SEDM, we cannot test this effect in this work.
However, it will be an interesting test with a larger sample of higher-resolution spectra in the future (e.g., $\sim$35,000 live transients spectra from 4MOST-TiDES with $R\sim6,500$) and testing with 4MOST-simulated spectra is ongoing (A. Milligan et al., in preparation).

There may also be some impact on the classification performance from having different templates in each tool.
\snid{} and \dash{} in this work share $\sim$93\% of their templates (4869 out of 5250 template spectra), because the \snid{} templates collected by \citetalias{Kim2022} were based on a training set for \dash{}.
Most of the \ngsf{} templates overlap with \snid{} templates, while the number of templates is lower (999 spectra from 187 objects) than other tools.
Therefore, we expect that there is no impact of different templates on classification between \snid{} and \dash{}, but there may be some impact between \ngsf{} and the other two classification tools.
In this work, we used the publicly available tools for accessibility.
We suggest that it would be beneficial to have consistent templates used in all spectral classification tools.

As noted in Sec.~\ref{subsec:eval_metrics}, the sample sizes of some types, mainly NonIa types, are small.
We need more spectra, especially for NonIa types, to calculate and use the performance evaluation metrics as more powerful supporting metrics for spectral classification.

In addition, updating current tools with recent data is required.
For \snid{} and \ngsf{}, more diverse templates across all types from SEDM and the Australian Dark Energy Survey \citep{Lidman2020} will help to classify transients more accurately and also to resolve ``type attractor'' issues.
For \dash{}, the model needs to be re-trained, for example, with the spectra available from the Weizmann Interactive Supernova Data Repository\footnote{\href{https://www.wiserep.org/}{https://www.wiserep.org/}} \citep[WISeREP;][]{Yaron2012} and to include SLSN and NotSN types to improve the overall performance evaluation metrics.

\ngsf{} shows the best performance among the three different tools, despite having fewer templates.
We believe that this is because \ngsf{} tries to separate SN light from host galaxy light by fitting them simultaneously. 
Implementing this technique into \snid{} and \dash{}\footnote{The current version of \dash{} can classify an SN and a host galaxy simultaneously only when the redshift is known.} would probably improve their classification accuracy.
On the other hand, more effort on separating between the SN and the host galaxy at the data level (e.g., a \contsep{} module in SEDM developed by \citetalias{Kim2022}) are required.

Another way to improve the classification accuracy is to use photometric information, such as light-curve shape and phase or host galaxy morphology and stellar mass.
Particularly, host galaxy information is known to be helpful in classifying transients, hence some studies are underway to implement this host information into photometric classification tools \citep{Gagliano2021, Gagliano2023}.
However, there are no such efforts on the spectral classification tool side.
Future spectroscopic surveys would benefit from the development of new tools that combine photometric information.

Lastly, one obvious way to improve the classification accuracy is to use a host galaxy redshift, ideally a spectroscopic one, in the classification process. 
The redshift data will be taken from publicly available galaxy catalogues, such as the Sloan Digital Sky Survey's extended Baryon Oscillation Spectroscopic Survey \citep{eboss} and the Dark Energy Spectroscopic Instrument's MOST Hosts Survey \citep{desi}.
Developing an efficient algorithm to implement this redshift information in the classification process is required.

Combining these various efforts will provide an accurate classification for newly discovered transients.
From this accurate classification, we can discover new types of astronomical objects and widen our understanding of the nature of those transients.
For SN Ia cosmology, we can estimate unbiased cosmological parameters without contamination from NonIa types in the LSST-era.

\section*{Acknowledgements}

We thank the anonymous referee for constructive suggestions and careful reading to clarify the manuscript.

Y.-L.K. acknowledges support from the Science and Technology Facilities Council [grant number ST/V000713/1]. 

I.H. is supported by the Science and Technology Facilities Council (STFC) research grants ST/V000713/1 and ST/Y001230/1 and from the Leverhulme Trust in the form of an International Fellowship, reference  IF-2023-027.

L.G. acknowledges financial support from AGAUR, CSIC, MCIN and AEI 10.13039/501100011033 under projects PID2020-115253GA-I00, PIE 20215AT016, CEX2020-001058-M, and 2021-SGR-01270.

U.B. and G.D. are supported by the H2020 European Research Council grant no. 758638.

T.E.M.B. acknowledges financial support from the Spanish Ministerio de Ciencia e Innovaci\'on (MCIN), the Agencia Estatal de Investigaci\'on (AEI) 10.13039/501100011033, and the European Union Next Generation EU/PRTR funds under the 2021 Juan de la Cierva program FJC2021-047124-I and the PID2020-115253GA-I00 HOSTFLOWS project, from Centro Superior de Investigaciones Cient\'ificas (CSIC) under the PIE project 20215AT016, and the program Unidad de Excelencia Mar\'ia de Maeztu CEX2020-001058-M.

Based on observations obtained with the Samuel Oschin Telescope 48-inch and the 60-inch Telescope at the Palomar Observatory as part of the Zwicky Transient Facility project. ZTF is supported by the National Science Foundation under Grants No. AST-1440341 and AST-2034437 and a collaboration including current partners Caltech, IPAC, the Weizmann Institute of Science, the Oskar Klein Center at Stockholm University, the University of Maryland, Deutsches Elektronen-Synchrotron and Humboldt University, the TANGO Consortium of Taiwan, the University of Wisconsin at Milwaukee, Trinity College Dublin, Lawrence Livermore National Laboratories, IN2P3, University of Warwick, Ruhr University Bochum, Northwestern University and former partners the University of Washington, Los Alamos National Laboratories, and Lawrence Berkeley National Laboratories. Operations are conducted by COO, IPAC, and UW.

SED Machine is based upon work supported by the National Science Foundation under Grant No. 1106171.

This analysis used \textsc{\texttt{pandas}} \citep{McKinney2010}, \textsc{\texttt{numpy}} \citep{Harris2020}, \textsc{\texttt{scipy}} \citep{Virtanen2020}, and \textsc{\texttt{matplotlib}} \citep{Hunter2007}.


\section*{Data Availability}

Performance evaluation metrics used for Fig.~\ref{fig:eval_metrics} are available on the \texttt{zenodo} webpage:  \href{https://doi.org/10.5281/zenodo.13144566}{https://doi.org/10.5281/zenodo.13144566}.

\snid{} templates used in this work are also provided upon request.



\bibliographystyle{mnras}




\appendix


\bsp	
\label{lastpage}
\end{document}